\documentclass[pra,twocolumn]{revtex4-1}
\usepackage{hyperref} 

\usepackage{graphicx} 
\usepackage{subfig}
\usepackage{color}
\usepackage{filecontents}
\usepackage{soul}
\usepackage{bbold}

\newcommand{\ket}[1]{\left| #1 \right\rangle}
\newcommand{\bra}[1]{\left\langle #1 \right|}

\newcommand{\be}{\begin{equation}}
\newcommand{\ee}{\end{equation}}
\newcommand{\bea}{\begin{eqnarray}}
\newcommand{\eea}{\end{eqnarray}}
\newcommand*{\myeqref}[2][Eq.~]{%
  \hyperref[{#2}]{#1(\ref*{#2})}%
}
\def\equationautorefname#1#2\null{%
  Eq.#1(#2\null)%
}

\usepackage{dcolumn}
\usepackage{bm}
\usepackage{amssymb,amsmath}
\usepackage{color}
\usepackage{float}
\usepackage{tikz}
\usetikzlibrary{arrows}

\definecolor{DarkGreen}{rgb}{0,0.6,0.2}

\begin{document}
\title{Coherent Perfect Absorption in Tavis-Cummings Models}
\author{Zibo Wang, Pawan Khatiwada, Dan Wang, and Imran M. Mirza}
\affiliation{Macklin  Quantum  Information Sciences, \\
Department of Physics, Miami University, Oxford, Ohio 45056, USA}
\email{mirzaim@miamioh.edu}


\begin{abstract}
We theoretically study the conditions under which two laser fields can undergo Coherent Perfect Absorption (CPA) when shined on a single-mode bi-directional optical cavity coupled with two two-level quantum emitters (natural atoms, artificial atoms, quantum dots, qubits, etc.). In addition to being indirectly coupled through the cavity-mediated field, in our Tavis-Cummings model the two quantum emitters (QEs) are allowed to interact directly via the dipole-dipole interaction (DDI). Under the mean-field approximation and low-excitation assumption, in this work, we particularly focus on the impact of DDI on the existence of CPA in the presence of decoherence mechanisms (spontaneous emission from the QEs and the leakage of photons from the cavity walls). We also present a dressed-state analysis of the problem to discuss the underlying physics related to the allowed polariton state transitions in the Jaynes-Tavis-Cummings ladder. As a key result, we find that in the strong-coupling regime of cavity quantum electrodynamics, the strong DDI and the emitter-cavity detuning can act together to achieve the CPA at two laser frequencies tunable by the inter-atomic separation which are not possible to attain with a single QE in the presence of detuning. Our CPA results are potentially applicable in building quantum memories that are an essential component in long-distance quantum networking.
\end{abstract}


\maketitle
\section{Introduction}
Coherent Perfect Absorption or CPA, in short, is an interference phenomenon in which the light fields incident on a lossy medium are neither reflected nor transmitted but completely absorbed by the medium through the process of destructive interference among all scattering amplitudes \cite{baranov2017coherent, baldacci2015coherent, yan2018coherent}. The captured light is then converted to other forms of energy inside the medium such as electrical energy or heat while the media in which CPA is achieved are called coherent perfect absorbers. The CPA is a material independent phenomenon and typically hinges on maintaining correct phases and amplitudes of the incident light beams. In principle, CPA is not confined to optical domains of wave-matter interactions and as a result, it has been reported to be attained with sound waves \cite{meng2017acoustic} and in diverse types of materials such as graphene films \cite{rao2014coherent}, Bose-Einstein condensates \cite{mullers2018coherent}, plasmonic metasurfaces \cite{zhang2012controlling}, and (broadly speaking) in the generic type of complex scattering structures with and without hidden symmetries \cite{chen2020perfect}, etc. 

Historically, Douglas Stone et al. \cite{chong2010coherent} were the first research group to put forward the idea of CPA in 2010. In their celebrated work, they introduced CPA as a time-reversed laser or anti-laser due to its ability to completely absorb the incident coherent fields with the proper adjustment of the phases and the intensities of the incoming light beams. In their theoretical proposal, they used a slab of Si material and incident light in the wavelength range $500-900~nm$. What they were able to show was that by adding the right amount of dissipation and controlling the phase and amplitude of the light beams accordingly, 99.9\% of the incident light can be absorbed at discrete frequency values. Within a year, in 2011 two important developments were made. First was Stefano Longhi's theoretical proposal of combining a laser and an anti-laser in a single optical device while fulfilling the condition of $\mathcal{P}\mathcal{T}$ symmetry \cite{longhi2010pt}. Secondly, the experimental demonstration of the CPA was reported by Wan et al. using a Ti-sapphire light source shined onto a 110$\mu m$ thick silicon wafer. They were able to achieve 99.4\% of absorption experimentally in the infrared wavelength regime \cite{wan2011time}. However, the actual proposal involving the Si slab suffered the limitation of only being workable in a narrowband regime. This restriction was relaxed in 2012 with the work of Pu et al. where they theoretically proposed perfect absorption in a broadband regime using heavily doped ultrathin Si films \cite{pu2012ultrathin}. Pu et al.'s work was experimentally verified by Li et al. in 2015 \cite{li2015broadband}. Advancing the possible applications of the CPA, in 2019 Pichler et al. demonstrated the operation of an anti-laser in random media \cite{pichler2019random} exhibiting the time-reversed version of random lasers \cite{wiersma2008physics}.

Fueled by the aforementioned notable developments, the CPA has shown several important applications in the past decade. Examples include photodetection instrumentation \cite{zheng2014broadband}, solar cells \cite{villinger2021doubling}, switches \cite{fang2014ultrafast}, microwave \& radiowave antennas \cite{grimm2021driving}, sensors \cite{kravets2013singular}, and logic operators \cite{papaioannou2016all}, to name a few. Entering the quantum domain, an immediate application of CPA turns out to be in the storage of quantum information that is encoded in different photonic degrees of freedoms \cite{lvovsky2009optical}. Such type of information storage is a vital step in performing typical protocols of quantum communication where flying photons are utilized as information carriers while QEs/qubits reside at the nodes of a network to store and manipulate the information \cite{kimble2008quantum, mirza2016multiqubit, mirza2016two, mirza2013single, mirza2015bi}. For reliable information processing the complete photon absorption by lossy qubits trapped inside leaky optical cavities becomes an indispensable requirement that can be fulfilled by CPA. In this regard, CPA has been investigated in cavity quantum electrodynamics (CQED) setups with a single two-level atom \cite{wei2018coherent}, single two-level atom and a second-order nonlinear crystal embedded in an optical cavity \cite{xiaong2020coherent}, collection of two-level atoms in the linear regime of CQED \cite{agarwal2015photon}, a two-level atom collection in the nonlinear regime of CQED with the possibility of optical bi-stability \cite{agarwal2016perfect,wang2017control}, three-level atomic ensemble CQED \cite{wang2017interference}, path entangled photons in Fabry-Per\'ot cavity with an absorbing medium \cite{huang2014coherent}, squeezed light beams incident upon absorbing beam splitters \cite{hardal2019quantum}, and in cavity quantum optomechanics \cite{yan2014coherent, zhang2016perfect}. 

\begin{figure*}
\centering
\includegraphics[width=5.2in, height=2.2in]{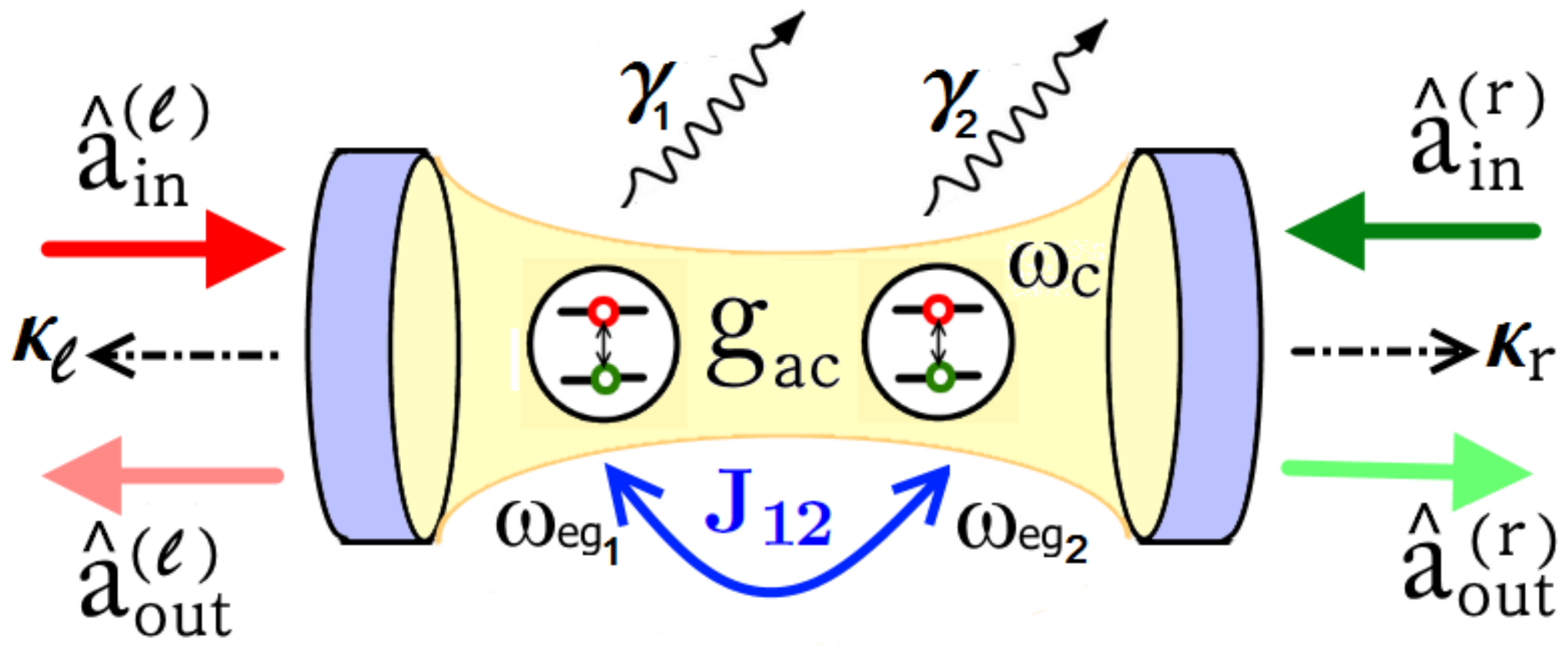} 
\captionsetup{
  format=plain,
  margin=1em,
  justification=raggedright,
  singlelinecheck=false
}
\caption{(Color online) The setup for CPA in the Tavis-Cummings model. Two identical QEs trapped inside a single-mode optical cavity are directly coupled to each other through DDI. Two incoming laser beams pump the cavity mode and we find the conditions under which the incoming fields are completely absorbed even in the presence of spontaneous emission from the emitters and photon losses from the imperfect cavity mirrors.}
\label{Fig1}
\end{figure*}

In all of the above-mentioned studies, the position of the QEs inside the cavity is completely ignored. In particular, for many-atom scenarios, a single atomic excitation operator is introduced which linearly sums up the excitation operator for each atom. Such type of treatment (considering an atomic ensemble as a single giant atom essentially) may work well in the standard Dicke regime \cite{dicke1954coherence} where the wavelength $\lambda$ of the resonant field is much larger than the size of the atomic ensemble $R$ i.e. $\lambda\gg R$. However, in the opposite situations ( $\lambda\lesssim R$; the so-called timed-Dicke regime \cite{scully2006directed, svidzinsky2009evolution}) the separation between the emitters cannot be neglected. Furthermore, it is known that once the inter-emitter separation $R$ becomes smaller compared to $\lambda$, quantum emitters tend to directly exchange the excitation through the dipole-dipole interaction \cite{agarwal2012quantum, mirza2020dimer}. 

Keeping in view this unexplored area of CPA in CQED, in this paper, we present an in-depth study of CPA with two QEs trapped inside an optical cavity with separation resolvable by the cavity mode field. Our system consists of two laser beams shined on the mirrors of a bidirectional single-mode optical cavity interacting with a pair of two-level QEs with dipole-dipole interaction (DDI). By deriving the quantum Langevin equations and input-output relations for atomic and field operators, we calculate the conditions under which the transmission and reflection of laser beams can be completely suppressed. We also performed a dressed state analysis to predict the frequencies at which the incident beams are completely absorbed to excite the polariton state transitions in the Jaynes-Tavis-Cummings ladder. As the main result of our investigation, we conclude that for an in-resonance emitter-cavity situation the strong DDI can diminish the CPA under the strong coupling regime of CQED. However, by introducing large detunings while still working in the strong DDI and strong CQED regimes, absorption can not only be enhanced by almost $40\%$ but CPA can be achieved at two distinct laser frequencies --- a feature that is not possible to realize with a single QE.

The paper is organized as follows. In Sec.~II we outline the details of our Tavis-Cummings model. In Sec.~III we discuss the dissipative dynamics of our setup through the quantum Langevin equations and input-output formalism. In Sec.~IV we report the conditions under which CPA can be observed. Therein we also report a dressed state analysis for the interpretation of the results which are discussed in Sec.~V. Finally, in Sec.~VI we close with a summary of our findings and mention the future outlook.


\section{System Model}
As shown in Fig.~\ref{Fig1}, we consider a bidirectional optical cavity with a single-mode resonant frequency $\omega_c$. The annihilation (creation) of a photon in the quantized optical mode is performed through the operator $\hat{a} (\hat{a}^\dagger)$. Under realistic conditions, photons can leak out from the imperfect cavity mirrors. We incorporate this process in our model by introducing the left (right) cavity leakage rate $\kappa_l (\kappa_r)$. We also consider two two-level QEs (quantum dots, superconducting Josephson junctions, naturally occurring Alkali atoms for instance Cs, etc.) trapped inside the optical cavity. We describe the process of emitter excitation (deexcitation) through the atomic raising (lowering) operators $\hat{\sigma}^\dagger_i(\hat{\sigma}_i)$ for the $i$th QE. In terms of the energy eigenstates of the bare QEs these atomic operators are defined as $\hat{\sigma}_i:=\ket{g_i}\bra{e_i}$ and $\hat{\sigma}^\dagger_i:=\ket{e_i}\bra{g_i}$ where $i=1,2$. Atomic transition frequency and spontaneous emission rate for the $1^{st}(2^{nd})$ QE is given by $\omega_{eg_{1}}$ and $\gamma_1$ ($\omega_{eg_{2}}$ and $\gamma_2$), respectively. Furthermore, each QE interacts with the single-mode field through the standard Jaynes-Cummings interaction with the coupling strength characterized by the parameter $g_{ac}$. For simplicity, we set the coupling strength to be the same for both QEs. Additionally, we suppose that the separation between the QEs is less than a wavelength which opens the possibility of direct coupling between QEs through the dipole-dipole interaction (DDI). The strength of the DDI $J_{12}$ sensitively depends on the inter-emitter separation and is defined as \cite{zhang2014effects}
\begin{align}\label{DDI}
J_{12}=\frac{3\Gamma_0 c^3}{4\omega^3_{eg}r^3_{12}}\left(1-3\cos^2\phi\right),
\end{align}
where $\Gamma_0$ is the atomic decay rate in the free space, $c$ is the speed of light, $r_{12}$ is the inter-emitter separation, and $\phi$ is the angle between the atomic dipole moment vector and the inter-emitter vector. For simplicity, in the definition of $J_{12}$ we have assumed both QEs to be identical i.e. $\omega_{eg_{1}}=\omega_{eg_{2}}=\omega_{eg}$, and $\Gamma_{0_1}=\Gamma_{0_2}=\Gamma_0$. Furthermore, we assume atomic dipole vectors to be perpendicular to the inter-emitter axis vector such that $\cos\phi=0$.

To study coherent perfect absorption (CPA) we consider that two input lasers with the same carrier frequency $\omega_l$ are being shined on our CQED setup. One laser enters the cavity from the left mirror and the other from the right mirror. Input operators $\hat{a}^{(l)}_{in}$ and $\hat{a}^{(r)}_{in}$ are used to describe the annihilation of photons in the left and right incoming lasers, respectively, while the corresponding output operators are given by $\hat{a}^{(l)}_{out}$ and $\hat{a}^{(r)}_{out}$. These operators are defined in the next section. 

Putting everything together, the system Hamiltonian $\hat{\mathcal{H}}_{sys}$, under the rotating wave approximation and in a frame rotating with the laser frequency $\omega_l$, is given by
\begin{align}\label{SysH}
\hat{\mathcal{H}}_{sys} = &~\hbar\Delta_c\hat{a}^{\dagger}\hat{a}+\sum_{i=1,2}\hbar\Delta_{eg_{i}}\hat{\sigma}^\dagger_i\hat{\sigma}_i+\sum_{i=1,2}\hbar\Big(g_i\hat{a}\hat{\sigma}^\dagger_i\nonumber\\
&+g^\ast_i\hat{\sigma}_i\hat{a}^\dagger\Big)+ \hbar J_{12}\left(\hat{\sigma}^\dagger_1\hat{\sigma}_2+\hat{\sigma}^\dagger_2\hat{\sigma}_1\right),
\end{align}
where $\Delta_c=\omega_c-\omega_l$, $\Delta_{eg_{i}}=\omega_{eg_{i}}-\omega_{l}$ are the cavity and QEs detunings, respectively. Non-vanishing commutation relations among different system operators are given by
\begin{align*}
\left[\hat{a},\hat{a}^\dagger\right]=1, \hspace{1mm}\text{and}\hspace{1mm} \{\hat{\sigma}^\dagger_j, \hat{\sigma}_k \}=\delta_{jk},\hspace{1mm}\text{with}~~j=1,2;~k=1,2.
\end{align*}
The first term on the right-hand side of Eq.~\eqref{SysH} describes the free Hamiltonian of the optical cavity. The second term represents the free Hamiltonian of the first and second QE. In the QE Hamiltonian, we have assumed the ground state energies to be zero. The third and fourth terms show the Jaynes-Cummings interaction between QEs and the single mode of the optical cavity. The fifth and sixth terms represent the DDI between the QEs.


\section{Dissipative System Dynamics and Output Fields}
Under the realistic treatment of the problem, we assume that the system is open to interact with the environmental degrees of freedom. Consequently, we decompose the total Hamiltonian $\hat{\mathcal{H}}$ of the global system (including QEs, optical cavity, and environment) into three pieces as
\begin{align}\label{TotH}
\hat{\mathcal{H}}=\hat{\mathcal{H}}_{sys}+\hat{\mathcal{H}}_{bath}+\hat{\mathcal{H}}_{int}.
\end{align}
The system Hamiltonian $\hat{\mathcal{H}}_{sys}$ has already been specified in Eq.~\eqref{SysH}. We focus on bath Hamiltonian $\hat{\mathcal{H}}_{bath}$ now. In there, to describe the cavity mode leakage from the imperfect cavity mirrors, we introduce two multimode quantum harmonic oscillators which model the environment(/bath/reservoir) Hamiltonian $\hat{\mathcal{H}}_{bath}$ for the intra-cavity field. One multimode environment for the left (or $l$) mirror and the other for the right (or $r$) mirror. The annihilation of a photon in the $\omega~ (\nu)$th mode of the left (right) bath is carried out by the operator $\hat{b}_l(\omega)\left(\hat{b}_r(\nu)\right)$. Moreover, to incorporate spontaneous emission from both QEs we assume two independent baths. We use $\hat{S}_1(\mu)$ and $\hat{S}_2(\xi)$ annihilation operators to describe the photon destruction in the $\mu$th and $\xi$th modes, respectively. With these considerations, the bath Hamiltonian takes the form
\begin{align}
&\hat{\mathcal{H}}_{bath} =~\hbar\int^{\infty}_{-\infty}\hspace{-2mm}\omega \hat{b}^\dagger_l(\omega)\hat{b}_l(\omega)d\omega + \hbar\int^\infty_{-\infty}\hspace{-2mm}\nu \hat{b}^\dagger_r(\nu)\hat{b}_r(\nu)d\nu \nonumber\\
&+ \hbar\int^\infty_{-\infty}\hspace{-2mm}\mu \hat{S}^\dagger_1(\mu)\hat{S}_1(\mu)d\mu +\hbar \int^\infty_{-\infty}\hspace{-2mm}\xi \hat{S}^\dagger_2(\xi)\hat{S}_2(\xi)d\xi,
\end{align}
where the non-vanishing commutation relations among different bath operators are given by 
\begin{align*}
&\left[\hat{b}_l(\omega), \hat{b}^\dagger_l(\omega^{'})\right] =\delta(\omega-\omega^{'}),~ \left[\hat{b}_r(\nu), \hat{b}^\dagger_r(\nu^{'})\right]=\delta(\nu-\nu^{'});\nonumber\\
& \left[\hat{S}_1(\mu), \hat{S}^\dagger_1(\mu^{'})\right]=\delta(\mu-\mu^{'}),~\left[\hat{S}_2(\xi), \hat{S}^\dagger_2(\xi^{'})\right]=\delta(\xi-\xi^{'}).
\end{align*}
Finally, the interaction Hamiltonian $\hat{\mathcal{H}}_{int}$, describing the coupling between the optical mode and QEs with their respective environmental degrees of freedom, is expressed as
\begin{align}
&\hat{\mathcal{H}}_{int}=~\hbar\sqrt{\frac{\kappa_l}{2\pi}}\int^\infty_{-\infty}\left(\hat{a}^\dagger\hat{b}_l(\omega)+\hat{b}^\dagger_l(\omega)\hat{a}\right)d\omega +\hbar \sqrt{\frac{\kappa_r}{2\pi}}\times\nonumber\\
&\int^\infty_{-\infty}\left(\hat{a}^\dagger\hat{b}_r(\nu)+\hat{b}^\dagger_r(\nu)\hat{a}\right)d\nu+\hbar\sqrt{\frac{\gamma_1}{\pi}}\int^\infty_{-\infty}\Big(\hat{\sigma}^\dagger_1\hat{S}_1(\mu)+\nonumber\\
&\hat{S}^\dagger_1(\mu)\hat{\sigma}_1\Big)d\mu+\hbar\sqrt{\frac{\gamma_2}{\pi}}\int^\infty_{-\infty}\left(\hat{\sigma}^\dagger_2\hat{S}_2(\xi)+\hat{S}^\dagger_2(\xi)\hat{\sigma}_2\right)d\xi.
\end{align}
In the above equation, the terms with prefactor $\sqrt{\kappa_l/2\pi}~(\text{or}~\sqrt{\kappa_r/2\pi})$ represent the coupling of the cavity mode field with the left (or right) environment leading to photon leakage from the cavity while the terms with the prefactor $\sqrt{\gamma_1/2\pi}~(\text{or}~\sqrt{\gamma_2/2\pi})$ show the interaction of first (or second) QE with its environment that will lead to spontaneous emission. Note that in the description of system-environment coupling, we have applied the rotating wave approximation such that within the global system the excitation number is always conserved. Furthermore, we have made use of the Markov approximation according to which all decay rates ($\kappa_l, \kappa_r,\gamma_1$ and $\gamma_2$) are assumed to be independent of their respective bath mode frequency.

Next, to describe the dynamics of the open quantum system under study, we formulate the problem in the Heisenberg picture. Therein (and as outlined in Appendix A) the dissipative dynamics can be described in terms of Heisenberg-Langevin equations (a set of coupled differential equations for the system operators incorporating decay processes and quantum noise from the environment). Since we are interested in the spectrum emitted by our CQED system, we perform a trace operation on both sides of Eq.~\eqref{HEOM}. As a result, we obtain the following equations of motion for the expectation value of operators
\begin{widetext}
\begin{subequations}\label{AvgHEOM}
\begin{align}
\partial_t\left<\hat{a}\right> &= -\left(i\Delta_c+\frac{\kappa_l+\kappa_r}{2}\right)\left<\hat{a}\right>-i\sum\limits^{2}_{j=1}g^\ast_j\left<\hat{\sigma}_j\right>-\sqrt{\kappa_l}\left<\hat{a}^{(l)}_{in}\right>-\sqrt{\kappa_r}\left<\hat{a}^{(r)}_{in}\right>, \label{QLE1}\\    
\partial_t\left<\hat{\sigma}_j\right> &= -\left(i\Delta_{eg_j}+\gamma_j\right)\left<\hat{\sigma}_j\right>+ig_j\left<\hat{\sigma}_{z,j}\hat{a}\right>+iJ_{jk}\left<\hat{\sigma}_{z,j}\hat{\sigma}_k\right>+\sqrt{2\gamma_j}\left<\hat{\sigma}_{z,j}\hat{\sigma}_{in,j}\right>, \label{QLE2}\\    
\partial_t\left<\hat{\sigma}_{z,j}\right> =& -4\gamma_j\left(\frac{\left<\hat{\sigma}_{z,j}\right>+1}{2}\right)-2i\left(g_j\left<\hat{a}\hat{\sigma}^\dagger_j\right>-g^\ast_j\left<\hat{\sigma}_j\hat{a}^\dagger\right>\right)-2iJ_{jk}\left(\left<\hat{\sigma}^\dagger_j\hat{\sigma}_k\right>-\left<\hat{\sigma}^\dagger_k\hat{\sigma}_j\right>\right)\nonumber\\
&-2\sqrt{2\gamma_j}\left(\left<\hat{\sigma}^\dagger_j\hat{\sigma}_{in,j}\right>+\left<\hat{\sigma}^{\dagger}_{in,j}\hat{\sigma}_j\right>\right),\label{QLE3}
\end{align}
\end{subequations}
\end{widetext}
where $j=1,2$; $k=1,2$ and $j\neq k$ with $J_{12}=J_{21}=J$ and $\hat{\sigma}_{z,j}\equiv [\hat{\sigma}^\dagger_j,\hat{\sigma}_j]$. To proceed with the analytic calculations we now apply three approximations that are central to this work. The first approximation we employ is called the ``mean-field approximation'' \cite{agarwal2010electromagnetically} according to which we replace the mean value of product of operators with the product of the mean value of each operator i.e. for any two arbitrary operators $\hat{o}_1$ and $\hat{o}_2$ we write $\left<\hat{o}_1\hat{o}_2\right>=\left<\hat{o}_1\right>\left<\hat{o}_2\right>$. In the second approximation, we assume a symmetric cavity i.e. we set $\kappa_l=\kappa_r=\kappa$. And in the third approximation we apply the ``weak-excitation assumption" \cite{li2014tunable, hu2015extended} according to which we assume that for the majority of the time (on the time scale set by the inverse of the atomic decay rate) both QEs stay in their ground state implying $\left<\hat{\sigma}_{z,j}\right>\approx -1$, $j=1,2$. With these assumptions, we take the steady-state conditions and solve Eq. set~\eqref{AvgHEOM} which gives us the following form of the intracavity field
\begin{widetext}
\begin{align}\label{Avga}
\left<\hat{a}\right>=&\frac{-\sqrt{\kappa}\left(\left<\hat{a}^{(l)}_{in}\right>+\left<\hat{a}^{(r)}_{in}\right>\right)}{i\Delta_c+\kappa+\left(\frac{g^\ast_1g_2\left(i\Delta_{eg_1}+\gamma_1\right)+g_1g^\ast_2\left(i\Delta_{eg_2}+\gamma_2\right)-i\left(\left|g_1\right|^2+\left|g_2\right|^2\right)J_{12}}{\left(i\Delta_{eg_1}+\gamma_1\right)\left(i\Delta_{eg_2}+\gamma_2\right)+J^2_{12}}\right)},\\
\implies 
\left<\hat{a}^{(l)}_{out}\right>= &\left<\hat{a}^{(l)}_{in}\right>-\frac{\kappa \left(\left<\hat{a}^{(l)}_{in}\right>+\left<\hat{a}^{(r)}_{in}\right>\right)}{i\Delta_c+\kappa+\left(\frac{g^\ast_1g_2\left(i\Delta_{eg_1}+\gamma_1\right)+g_1g^\ast_2\left(i\Delta_{eg_2}+\gamma_2\right)-i\left(\left|g_1\right|^2+\left|g_2\right|^2\right)J_{12}}{\left(i\Delta_{eg_1}+\gamma_1\right)\left(i\Delta_{eg_2}+\gamma_2\right)+J^2_{12}}\right)}\label{Avgaoutl},
\end{align}
\end{widetext}
while writing the expectation value of the output field we have taken the trace of the input-output relations (as derived in Appendix B) i.e.
\begin{subequations}\label{AvgInOut}
\begin{align}
\left<\hat{a}^{(l)}_{out}\right> =& \left<\hat{a}^{(l)}_{in}\right>+\sqrt{\kappa_l}\left<\hat{a}\right>,\\
\left<\hat{a}^{(r)}_{out}\right> =& \left<\hat{a}^{(r)}_{in}\right>+\sqrt{\kappa_r}\left<\hat{a}\right>,
\end{align}
\end{subequations}
and inserted Eq.~\eqref{Avga} into Eq.~\eqref{AvgInOut}. Note that in Eq.~\eqref{Avgaoutl} we have neglected any direct input to the QEs i.e. we have set $\left<\hat{\sigma}_{in,j}\right>=0$. We notice that for the case of a single QE the above expression for $\left<\hat{a}\right>$ simplifies to the following result which is already reported by Agarwal et al. in Ref.~\cite{agarwal2016perfect}
\begin{align}\label{1QE}
\left<\hat{a}\right>=\frac{-\sqrt{\kappa}\left(\left<\hat{a}^{(l)}_{in}\right>+\left<\hat{a}^{(r)}_{in}\right>\right)}{i\Delta_c+\kappa+\frac{|g|^2}{(i\Delta_{eg}+\gamma)}}.   
\end{align} 
Since we have assumed identical inputs $\left<\hat{a}^{(r)}_{in}\right>=\left<\hat{a}^{(l)}_{in}\right>$ and intra-cavity parameters are identical for both left and right channels, therefore, we conclude $\left<\hat{a}^{(r)}_{out}\right>=\left<\hat{a}^{(l)}_{out}\right>$. Equipped with these mathematical results, in the next subsection, we concentrate on the conditions under which the CPA can be achieved.


\section{CPA conditions and the dressed State Analysis}
\subsection{CPA conditions}
The CPA requires all input fields to be completely absorbed resulting in no output field detection in both left and right channels i.e. $\left<\hat{a}^{(l)}_{out}\right>=\left<\hat{a}^{(r)}_{out}\right>=0$. Therefore, we impose this condition on the output fields found in the last section (Eq.~\eqref{Avgaoutl}). After some rearrangement of terms, we arrive at the following two conditions necessary to achieve CPA in the present setup
\begin{subequations}\label{CPAcond}
\begin{align}
&\kappa\left(-\Delta^2_{eg}+\gamma^2+J^2\right) =-2\Delta_c\Delta_{eg}\gamma+2g^2\gamma,\\
&2\kappa\gamma\Delta_{eg} =~\Delta_c\left(-\Delta^2_{eg}+\gamma^2+J^2\right)+2g^2\left(\Delta_{eg}-J\right).\
\end{align}
\end{subequations}
Here for the sake of simplicity we have assumed both QEs as well as both input fields to be identical i.e. we have set  $\Delta_{eg_1}=\Delta_{eg_2}=\Delta_{eg}$, $\gamma_1=\gamma_2=\gamma$, $g_1=g_2=g$ (with real-valued emitter-cavity coupling parameters), and  $\left<\hat{a}^{(l)}_{in}\right>=\left<\hat{a}^{(r)}_{in}\right>=\left<\hat{a}_{in}\right>$. We now develop a physical understanding of these conditions. To this end, we present two different descriptions. In the first description, we rely on the mathematical results and analyze the phase relationship among the incident and cavity-emitted light fields to study the destructive interference. In the second approach, we will be paying attention to the polariton state formation and inter-state transitions by performing a dressed-state analysis.

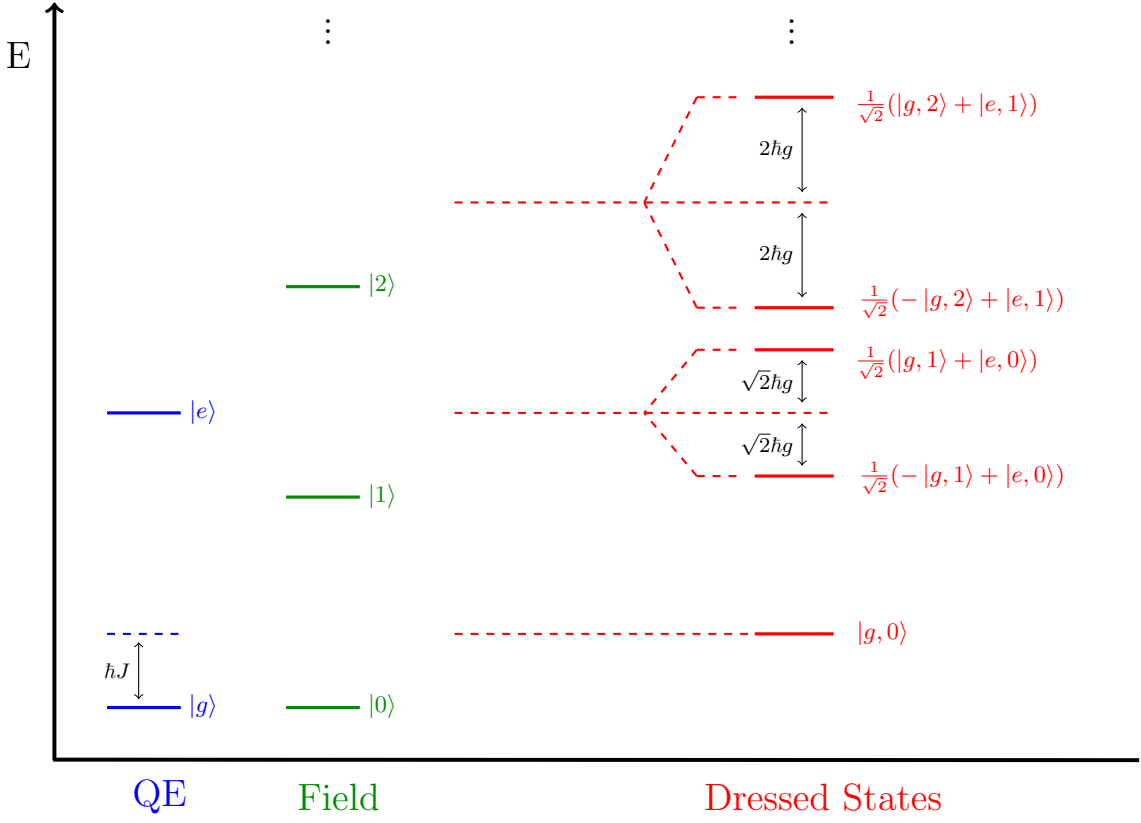
\begin{figure*}
\centering
\begin{tikzpicture}[scale=1.4]
\draw[, thick, ->,line width=1.5pt] (-0.3,-0.2) -- (-0.3,7);
\draw[-, thick,line width=1.5pt] (-0.312,-0.195) -- (10,-0.2);
\node at (-0.64, 6.5) {\Large E};
\draw[color=blue!, thick, dashed] (0.2,1) -- (0.9,1);
\draw[color=blue!, very thick] (0.2,3.1) -- (0.9,3.1)  node[label={[shift={(0.3,-0.4)}]$\ket{e}$}]{};
\draw[color=blue!, very thick] (0.2,0.3) -- (0.9,0.3)  node[label={[shift={(0.3,-0.4)}]$\ket{g}$}]{};
\draw[<->] (0.5 ,0.38) -- (0.5 ,0.92) node[midway, left] {\footnotesize $\hbar{J}$};
\node[color=blue!] at (0.7,-0.55) {\Large QE};
\draw [color=green!55!black!100, very thick](1.9,0.3) -- (2.6,0.3)  node[label={[shift={(0.3,-0.4)}]$\ket{0}$}]{};
\draw [color=green!55!black!100, very thick](1.9,2.3) -- (2.6,2.3)  node[label={[shift={(0.3,-0.4)}]$\ket{1}$}]{};
\draw [color=green!55!black!100, very thick](1.9,4.3) -- (2.6,4.3)  node[label={[shift={(0.3,-0.4)}]$\ket{2}$}]{};
\node[color=green!55!black!100] at (2.4,-0.55) {\Large Field};
\node at (2.3,6.8) {\textbf{\vdots}};
\draw [dashed, color=red!, thick] (3.5,1) -- (7.1,1);
\draw [color=red!, very thick] (6.35,1) -- (7.1,1)
node[label={[shift={(.65,-0.4)}]\small $\ket{g,0}$}]{};
\draw [color=red!, very thick] (6.35,3.7) -- (7.1,3.7)  node[label={[shift={(1.5,-0.65)}]\small $\frac{1}{\sqrt{2}}(\ket{g,1}+\ket{e,0})$}]{};
\draw [color=red!, very thick] (6.35,2.5) -- (7.1,2.5)  node[label={[shift={(1.7,-0.5)}]\small $\frac{1}{\sqrt{2}}(-\ket{g,1}+\ket{e,0})$}]{};

\node[color=red!] at (7,-0.55) {\Large{Dressed States}};
\node at (6.7,6.8) {\textbf{\vdots}};
        
\draw[dashed, color=red!, thick] (3.5,3.1) -- (7.1,3.1);
\draw[dashed, color=red!, thick] (5.8,3.7) -- (5.3,3.1);
\draw[dashed, color=red!, thick] (5.8,2.5) -- (5.3,3.1);
\draw[dashed, color=red!, thick] (5.8,3.7) -- (6.25,3.7);
\draw[dashed, color=red!, thick] (5.8,2.5) -- (6.25,2.5);
        
split 2
\draw[<->] (6.8,3.6) -- (6.8,3.2) node[midway, left] {\footnotesize $\sqrt{2}\hbar{g}$};
\draw[<->] (6.8,2.6) -- (6.8,3)node[midway, left] {\footnotesize $\sqrt{2}\hbar{g}$};

\draw [color=red!, very thick] (6.35,6.1) -- (7.1,6.1)  node[label={[shift={(1.5,-0.6)}]\small $\frac{1}{\sqrt{2}}(\ket{g,2}+\ket{e,1})$}]{};
\draw [color=red!, very thick] (6.35,4.1) -- (7.1,4.1)  node[label={[shift={(1.7,-0.4)}]\small $\frac{1}{\sqrt{2}}(-\ket{g,2}+\ket{e,1})$}]{};
        
\draw[dashed, color=red!, thick] (3.5,5.1) -- (7.1,5.1);
\draw[dashed, color=red!, thick] (5.8,6.1) -- (5.3,5.1);
\draw[dashed, color=red!, thick] (5.8,4.1) -- (5.3,5.1);
\draw[dashed, color=red!, thick] (5.8,6.1) -- (6.25,6.1);
\draw[dashed, color=red!, thick] (5.8,4.1) -- (6.25,4.1);
        
\draw[<->] (6.8,6.0) -- (6.8,5.2) node[midway, left] {\footnotesize $2\hbar{g}$};
\draw[<->] (6.8,4.2) -- (6.8,5.0) node[midway, left] {\footnotesize $2\hbar{g}$};
\end{tikzpicture}
\captionsetup{
format=plain,
margin=1em,
justification=raggedright,
singlelinecheck=false
}
\caption{(Color online) Energy-level diagram for two-QEs CQED architecture with DDI between QEs. A polariton state manifold with up to two excitations is drawn. In this diagram we have assumed an on-resonance case i.e. $\omega_{eg}=\omega_c$.}
\label{Fig2}
\end{figure*}


\subsubsection{Interpretation based on the CPA conditions}
To understand what is causing no filed detection at the two output channels, we first consider a single input situation and assume there is no input field present from the right channel of the cavity i.e. $\left<\hat{a}^{(r)}_{in}\right>=0$. Under these conditions, Eq.~\eqref{Avgaoutl}, and Eq.~\eqref{CPAcond} yield
\begin{align}
\left<\hat{a}^{(r)}_{out}\right> = -\frac{1}{2}\left<\hat{a}^{(l)}_{in}\right>, \hspace{5mm}\text{and}\hspace{5mm}\left<\hat{a}^{(l)}_{out}\right> = +\frac{1}{2}\left<\hat{a}^{(l)}_{in}\right>.
\end{align}
We notice that the expectation values of both outputs have the same magnitude but the right (left) channel output is out (in) phase from the input field by $\pi$ radian. Next, we assume the opposite situation and consider no input field from the left channel i.e. $\left<\hat{a}^{(l)}_{in}\right>=0$. In this case, we find
\begin{align}
\left<\hat{a}^{(r)}_{out}\right> = +\frac{1}{2}\left<\hat{a}^{(r)}_{in}\right>, \hspace{5mm}\text{and}\hspace{5mm}\left<\hat{a}^{(l)}_{out}\right> = -\frac{1}{2}\left<\hat{a}^{(r)}_{in}\right>.
\end{align}
We notice in this case the left output field becomes out of phase from the input field. Next, if we combine both of these situations and suppose identical input fields i.e.$\left<\hat{a}^{(l)}_{in}\right>=\left<\hat{a}^{(r)}_{in}\right>$ then we notice that both output fields vanish. We observe that once the CPA conditions are met there is a complete cancellation of the output fields in both left and right channels due to the destructive interference between the incident laser beams and the cavity-emitted field. It is also important to emphasize the need of using two identical input fields for such a perfect cancellation of fields at the output detection channels. We point out here that a similar interpretation has also been presented in Ref.~\cite{agarwal2015photon} in the context of perfect photon absorption in a single QE CQED setting. This shows that even though the DDI causes the intra-cavity field to change as compared to the intra-cavity field found in the single QE scenario but as long as the CPA conditions (which takes care of the DDI) are fulfilled the destructive interference still occurs at the output channels. 


\subsubsection{Dressed state analysis}
In order to study the impact of DDI on the energy level configuration of our Tavis-Cummings setup, in this subsection, we perform a dressed state analysis of the problem. It turns out that with the presence of the second QE trapped inside the optical cavity and directly interacting with the first QE through the DDI, it becomes rather difficult to calculate the dressed states of the combined atom-field system \cite{zhang2014effects}. However, this issue can be tackled under the weak-excitation approximation whereby following Ref.~\cite{nicolosi2004dissipation} we introduce a unitary transformation 
\begin{align}
\hat{U}=\exp\left[-\frac{\pi}{4}\left(\hat{\sigma}^\dagger_1\hat{\sigma}_2-\hat{\sigma}^\dagger_2\hat{\sigma}_1\right)\right].
\end{align}
Such a transformation can be used to reduce our two-QE CQED problem into an effective single QE CQED problem, as we show below. To this end, we apply the aforementioned $\hat{U}$ onto our original two QE system Hamiltonian (given in Eq.~\eqref{SysH}) and as a result, we obtain the following Hamiltonian 
\begin{align}\label{HamTrans}
\widetilde{H}_{sys}&\equiv\hat{U}^\dagger\hat{\mathcal{H}}_{sys}\hat{U}\nonumber\\
&=\hbar\Delta_{c}\hat{a}^\dagger\hat{a}+\hbar\left(\Delta_{eg}+J\right)\hat{\sigma}^\dagger_1\hat{\sigma}_1+\hbar\left(\Delta_{eg}-J\right)\hat{\sigma}^\dagger_2\hat{\sigma}_2\nonumber\\
&+\sqrt{2}\hbar g\left(\hat{a}\hat{\sigma}^\dagger_1+\hat{\sigma}_1\hat{a}^\dagger\right),
\end{align}
where we have adopted the simplification that both QEs have identical transition frequencies. We notice that in the transformed frame only the first QE (with a shifted frequency $\Delta_{eg}+J$) couples with the cavity field mode with a modified coupling strength $\sqrt{2}g$. The second QE (with a modified frequency $\Delta_{eg}-J)$) is completely decoupled from the field and hence can be neglected for focusing on the interaction part of the processes. Introducing a simplified notation in which $\hat{\sigma}_1=\hat{\sigma}$, $\hat{\sigma}^\dagger_1=\hat{\sigma}^\dagger$ and using the basis set $\lbrace \ket{e}\otimes\ket{n-1}\equiv\ket{e,n-1}$, $\ket{g}\otimes\ket{n}\equiv \ket{g,n}\rbrace$, one can then diagonalize the transformed Hamiltonian Eq.~\eqref{HamTrans}. Accordingly, we obtain the following eigensystem (with eigenvalues $\lambda_{\pm}$ and corresponding eigenvectors $\ket{\lambda_{\pm}}$)
\begin{align}
&\lambda_{\pm}=n\hbar\Delta_c+\frac{\hbar}{2}\widetilde{\Delta}_{ac}\pm\frac{\hbar}{2}\widetilde{\Omega}_n\left(\widetilde{\Delta}_{ac}\right),\label{Eigenpm}\\
&\ket{\lambda_+}=\cos\left(\frac{\Phi_n}{2}\right)\ket{e,n-1}+\sin\left(\frac{\Phi_n}{2}\right)\ket{g,n},\\
&\ket{\lambda_-}=-\cos\left(\frac{\Phi_n}{2}\right)\ket{g,n}+\sin\left(\frac{\Phi_n}{2}\right)\ket{e,n-1},
\end{align}
where $\widetilde{\Delta}_{ac}:=\Delta_{eg}-\Delta_c+J$, $\widetilde{\Omega}_n\left(\widetilde{\Delta}_{ac}\right):=\sqrt{\widetilde{\Delta}^2_{ac}+8g^2n}$, and $\tan\left(\frac{\Phi_n}{2}\right):=\frac{2g\sqrt{2n}}{\widetilde{\Delta}_{ac}}$. Note that the above results readily represent the single QE case as well if we set $J=0$ and replace $\sqrt{2}g\rightarrow g$. In Fig.~\ref{Fig2}, we have plotted the energy-level diagram of our Tavis-Cummings model. In the bare state situation (blue and green colored energy level) we have $2n$ number of uncoupled states possible (where $n$ being the cavity photon number). When the QE and the cavity field are allowed to interact, atom-cavity entanglement causes the polariton state formation (red-colored states shown in Fig.~\ref{Fig2}). When we compare two-QEs dressed states to the single QE problem, we observe that one of the main changes is that the excited dressed states are now split by an amount $2g\sqrt{2n}$. Furthermore, the ground state of the field-coupled single-QE is shifted by an amount $\hbar J$.


\section{Results and Discussion}
\begin{figure}
\centering
\includegraphics[width=3.4in, height=1.85in]{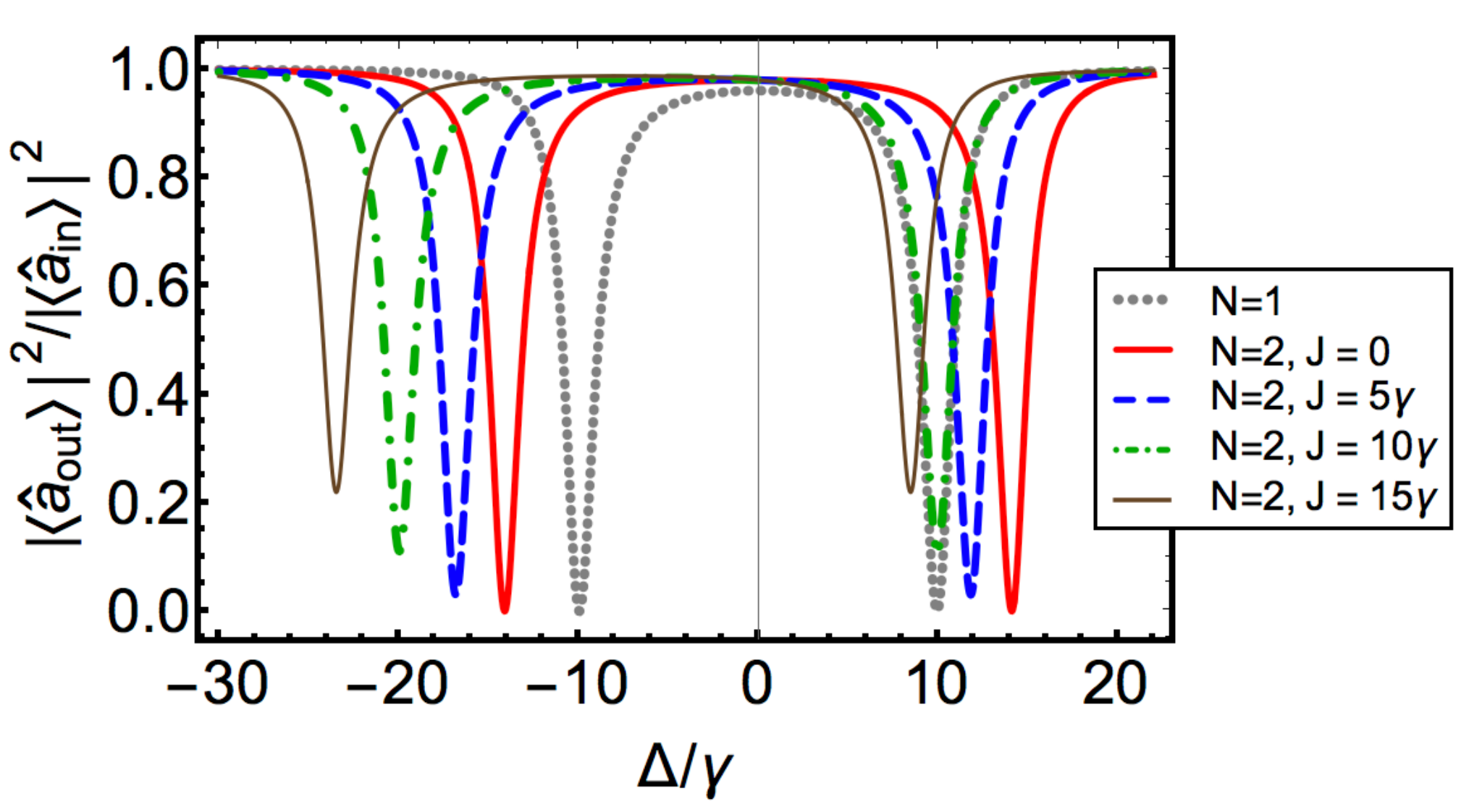}
\captionsetup{
format=plain,
margin=0.1em,
justification=raggedright,
singlelinecheck=false
}
\caption{The normalized output spectra as a function of detunning $\Delta=\omega_c-\omega_l$ for the case of two identical input fields (i.e. $\hat{a}^{(l)}_{in}=\hat{a}^{(r)}_{in}$ implying $\hat{a}^{(l)}_{out}$=$\hat{a}^{(r)}_{out}=\hat{a}_{out}$). In this figure we have assumed an on-resonance condition i.e. $\omega_{eg_1}=\omega_{eg_2}=\omega_{c}$ and varied the strength of the DDI $J_{12}=J$ between the two QEs ($N=2$ cases). For comparison $N=1$ (single QE) case is also displayed. The rest of the parameters are $\gamma_1=\gamma_2=\gamma$, $\kappa=\gamma$, $g_1=g_2=10\gamma$.}\label{Fig3}
\end{figure}

To set the stage of results, in Fig.~\ref{Fig3}, we plot the output spectra $\left|\left<{\hat{a}_{out}}\right>\right|^2/\left|\left<{\hat{a}_{in}}\right>\right|^2$ (also called the normal-mode spectra) in the presence of two identical inputs (and consequently identical outputs) as a function of the detuning $\Delta\equiv\omega_c-\omega_l$ with $\omega_{eg_j}=\omega_c, \forall j=1,2$. The main focus of this plot is to analyze the impact of the DDI on the spectral properties of our Tavis-Cummings setup. The results focusing on CPA conditions are presented in the next figure onward. For comparison, we have plotted the single QE ($N=1$) case (grey dotted curve) in Fig.~\ref{Fig3} as well. Using Eq.\eqref{1QE}, for $N=1$ case, we obtain
\begin{align}\label{1QEout}
\left<\hat{a}_{out}\right>=\left<\hat{a}_{in}\right>-\frac{2\kappa\left<\hat{a}_{in}\right>}{i\Delta+\kappa+\frac{|g|^2}{(i\Delta_{eg}+\gamma)}}.  
\end{align}
For the single QE case, we note that the spectrum is a frequency doublet with equal depths of both resonances. Moreover, the position of the resonances is symmetric around $\Delta=0$. Ignoring the spontaneous emission and cavity losses (which have been neglected while performing the dressed-state analysis), the separation between the two dips is given by $2g$. This is the typical Rabi splitting encountered in the Jaynes-Cummings model and it is regarded as the hallmark of the strong coupling regime of light-matter interaction in the CQED and circuit QED platforms \cite{thompson1998nonlinear,cui2006emission, wallraff2004strong}.
\begin{figure*}
\centering 
\includegraphics[width=2.275in, height=1.65in]{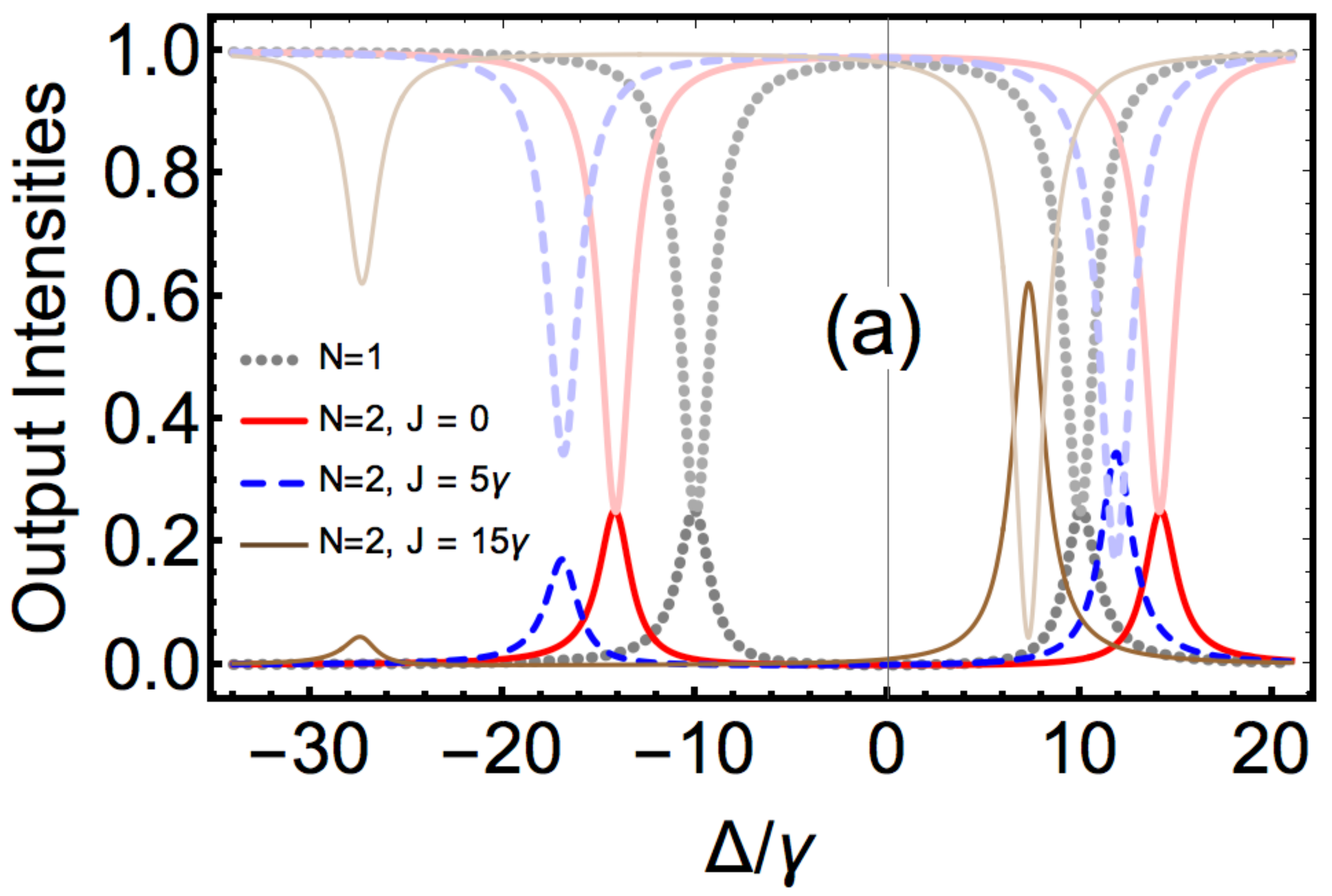}
\includegraphics[width=2.275in, height=1.65in]{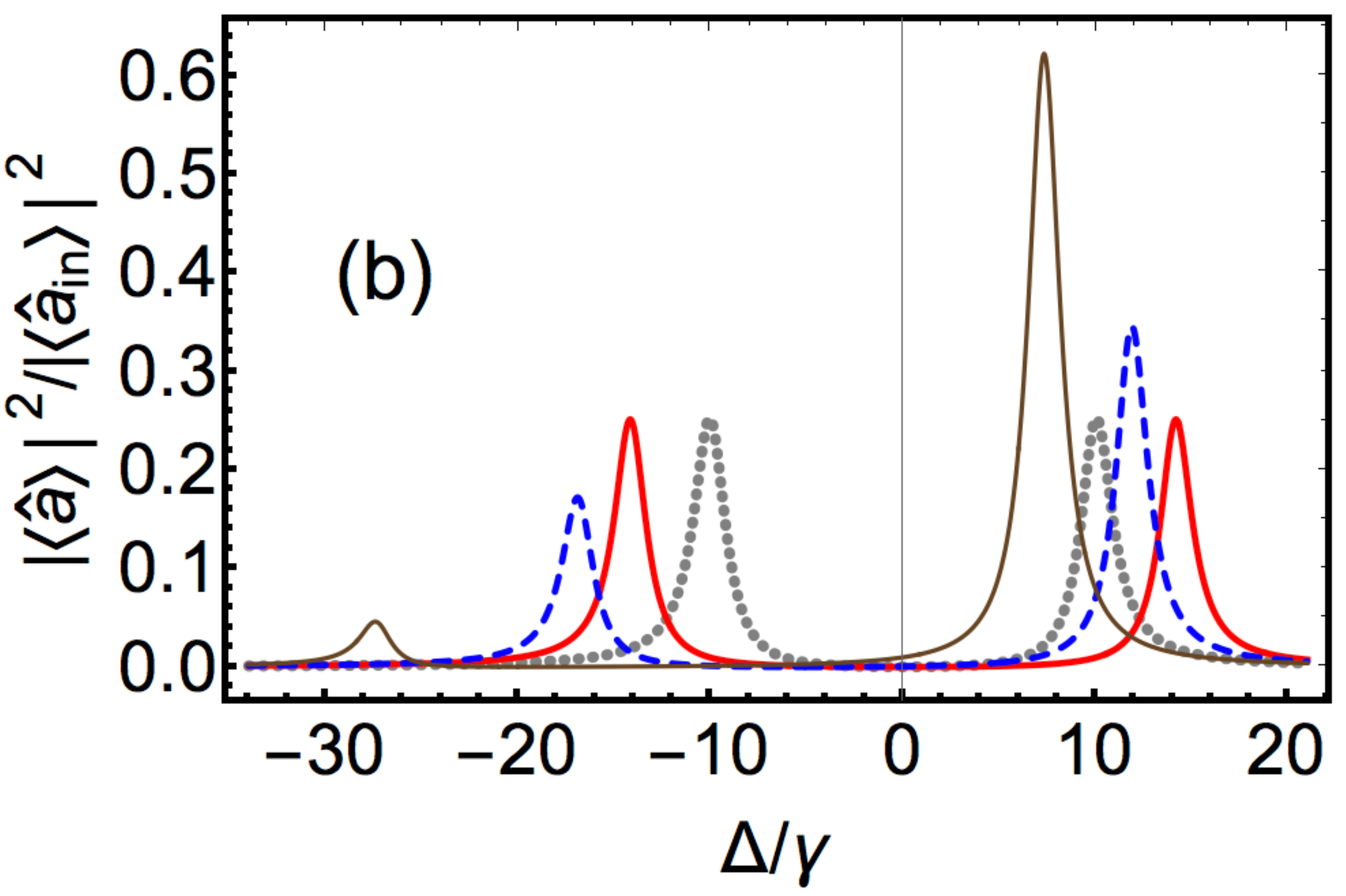}
\includegraphics[width=2.275in, height=1.65in]{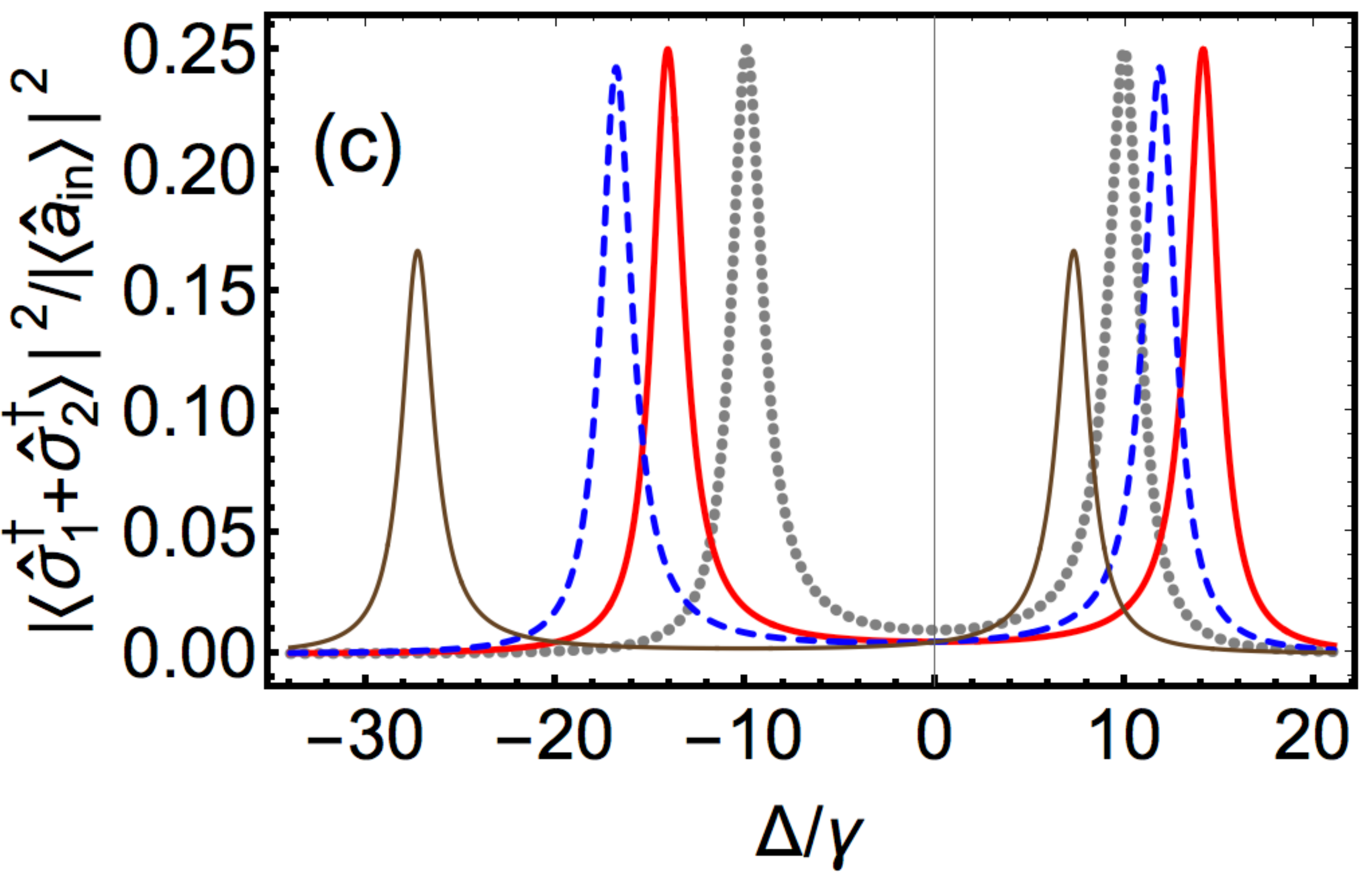}
\captionsetup{
format=plain,
margin=0.1em,
justification=raggedright,
singlelinecheck=false
}
\caption{The output field intensities (plot {\bf(a)}), the intracavity field (plot {\bf(b)}), and the collective emitter excitation probability (plot {\bf(c)}) for the case of a single input drive entering the cavity from the left hand side i.e. $\left<\hat{a}^{(r)}_{in}\right>=0$. In plot {\bf(a)}, the lighter (darker) colored curves are the output field intensities from the left (right) channel. The parameters are the same as used in the previous figure.}\label{Fig4}
\end{figure*}

\begin{figure*}
\centering 
\includegraphics[width=2.275in, height=1.65in]{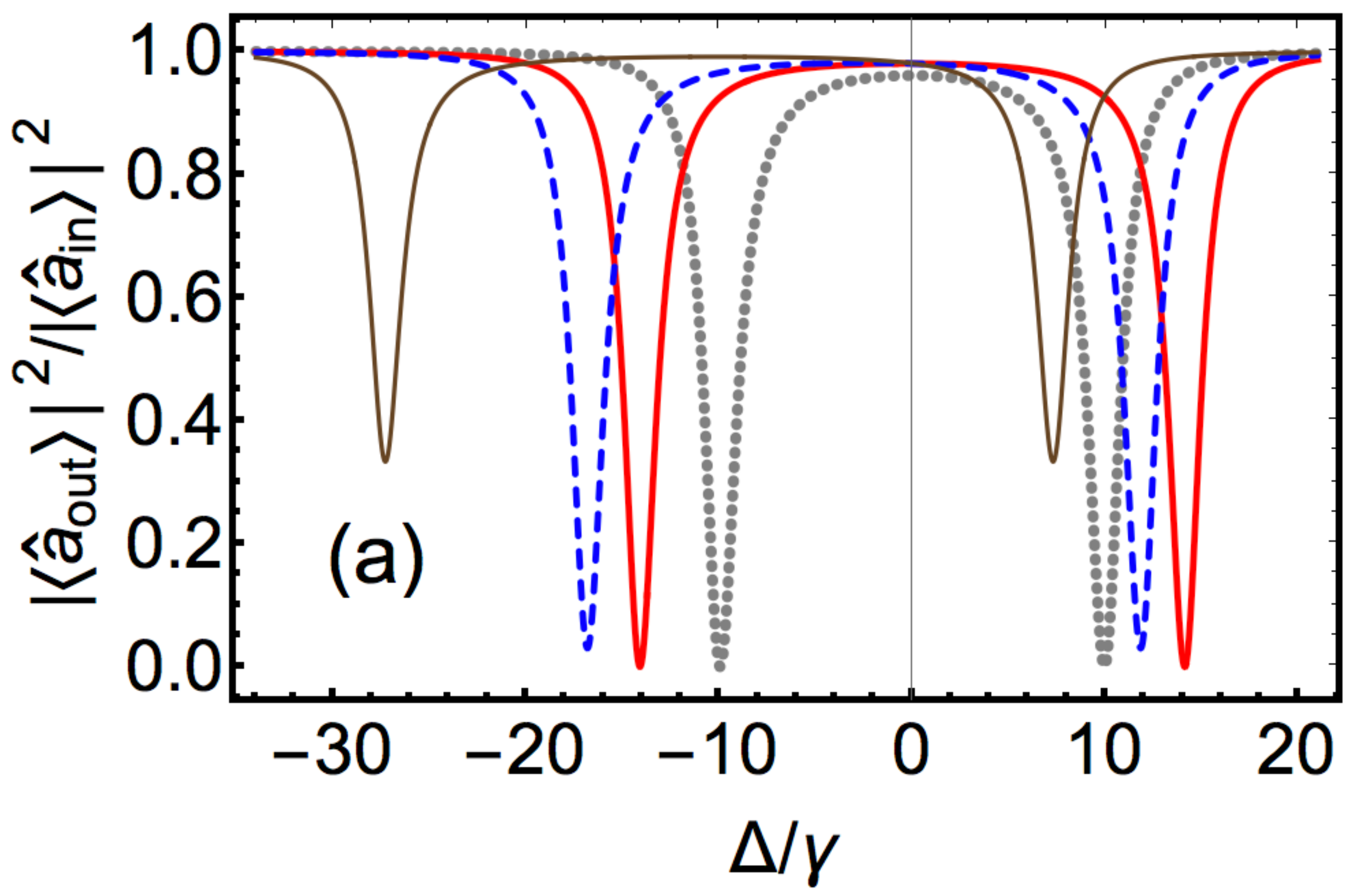}
\includegraphics[width=2.275in, height=1.65in]{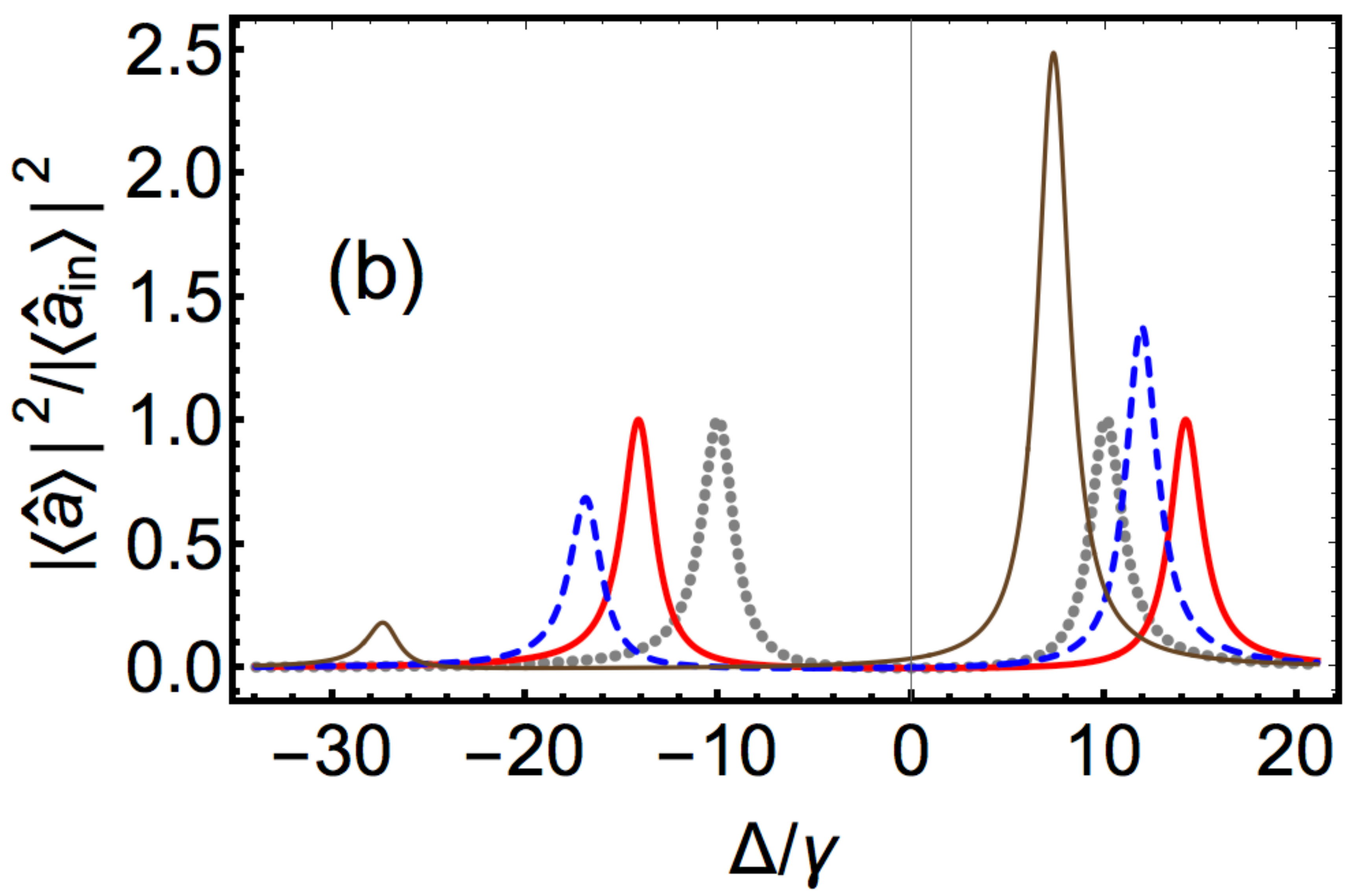}
\includegraphics[width=2.275in, height=1.65in]{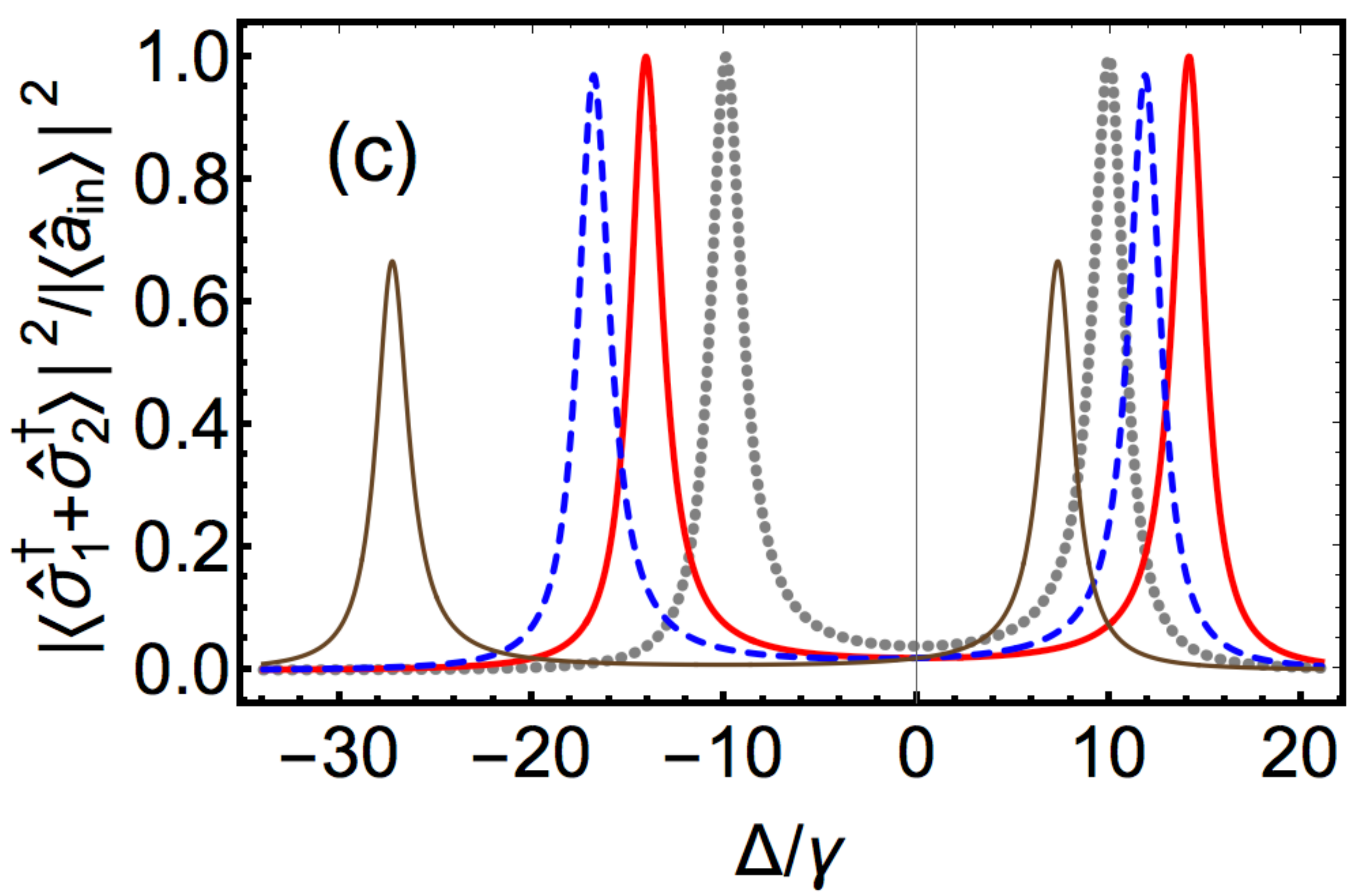}
\captionsetup{
  format=plain,
  margin=0.1em,
  justification=raggedright,
  singlelinecheck=false
}
\caption{{\bf(a)} Output intensities, {\bf(b)} intracavity field intensity, and {\bf(c)} collective emitter excitation probability plotted as a function of detuning $\Delta/\gamma$. Compared to Fig.~\ref{Fig4} the difference now is that we have assumed two identical input fields pumping the cavity mode from the left and right mirrors. The rest of the parameters and the curve types are the same as used in Fig.~\ref{Fig4}.}\label{Fig5}
\end{figure*}

Moving on to the two identical QE problem, we notice that the expectation value of the output operator takes the value
\begin{align}\label{AvgoutUp}
\left<\hat{a}_{out}\right>=\left<\hat{a}_{in}\right>-\frac{2\kappa \left<\hat{a}_{in}\right>}{i\Delta+\kappa+\left(\frac{2|g|^2 \left(i\Delta+\gamma\right)-2i\left|g\right|^2 J}{\left(i\Delta+\gamma\right)^2+J^2}\right)}.
\end{align}
As shown in Fig.~\ref{Fig3}, we note that for all cases of two QEs, the spectrum remains a frequency doublet. However, compared to the single QE case, the separation between the two dips, for $J=0$ case, takes the value $2\sqrt{2}g$. See the red thick solid curve with $g=10\gamma$, $J=0$ and we are observing a splitting of $\sim 28.2\gamma$. This observation is consistent with the prediction of the dressed state analysis where the peak separation (for either $J=0$ or $J\neq 0$ scenario) is expected to take the value $\Delta\lambda=\lambda_+ - \lambda_-=\hbar\widetilde{\Omega}_n\left(\widetilde{\Delta}_{ac}\right)$. Next, we notice that the separation between the resonances tends to increase as DDI $J$ is introduced and consequently enhanced (see blue dashed, green dotted dashed, and brown solid curves). This trend can again be explained using the dressed-state picture, where we expect the peak separation to increases monotonically with $J$ (i.e. $\Delta\lambda \propto \sqrt{8g^2+J^2}$). Additionally, we remark that the minima in the output spectra take a higher value as we increase $J$. This trend can be explained by calculating the transition matrix element  $\sqrt{\kappa}\bra{\lambda_{+}}\hat{a}\ket{g,0}=\sqrt{\kappa}\sin\left(\Phi_0/2\right)$ and $\sqrt{\kappa}\bra{\lambda_{-}}\hat{a}\ket{g,0}=-\sqrt{\kappa}\cos\left(\Phi_0/2\right)$. When $J=0$, it turns out that the QEs and cavity mode equally contribute such that the peak depths take the same value, where when $J\neq 0$ is introduced this equal contribution is disturbed. Overall in Fig.~\ref{Fig3}, we observe that the inclusion of two QEs allows us better control of the spectral properties by tuning the inter-emitter separation (which is equivalent to altering the DDI as from Eq.~\eqref{DDI} the inter-emitter separation $r_{12}$ is linked to the DDI $J$ through $r^3_{12}=\frac{3\Gamma_0 c^3}{4\pi^3 \omega^3_{eg}}\frac{1}{J}$).

Afterward, we focus on the perfect absorption of the input fields. In Fig.~\ref{Fig4} we consider the case of a single input drive from the left cavity channel (i.e. we set $\left<\hat{a}^{(r)}_{in}\right>=0$). In all plots of this figure, we are working in the strong coupling regime of CQED (with cooperativity $\mathcal{C}\equiv g^2/(2\kappa\gamma)=50$) where the single input drive is on resonance with the polariton states frequencies $\lambda_{\pm}\propto \widetilde{\Omega}_{n}/2$. As we have seen in Sec. IV A, for this case, the single input field fails to undergo complete destructive interference at the output channels $\Big(\left<\hat{a}^{(r)}_{out}\right>=-1/2\left<\hat{a}^{(l)}_{in}\right>=$ with $\left<\hat{a}^{(l)}_{out}\right>=+1/2\left<\hat{a}^{(l)}_{in}\right>\Big)$ and hence no CPA is achieved. Fig.~\ref{Fig4}(a) confirms this result for both single and double QE scenarios. For two QE cases, we notice as the DDI is enhanced the left output field intensities show a decreasing tendency on the positive $\Delta$-axis such that for $J=15\gamma$ case the left output field shows an almost complete suppression at $\Delta\approx 7.8\gamma$. However, CPA is still not achieved as at the same frequency, we notice the right output field reaches more than $60\%$ intensity value. Furthermore, on the negative $\Delta$-axis, the output field intensities show a trend opposite to their positive $\Delta$-axis behavior as $J$ is increased. 
In Fig.~\ref{Fig4}(b) and \ref{Fig4}(c) we present the intracavity field (photonic excitation) and total emitter excitation normalized to the input field $\hat{a}^{(l)}_{in}\equiv\hat{a}_{in}$, respectively. In both of these plots, we again observe a frequency doublet profile due to the strong coupling regime ($\mathcal{C}>1)$ of CQED. As a prominent feature, we observe that as $J>\kappa$ and eventually when $J>g$, the peak on the positive $\Delta$-axis of Fig.~\ref{Fig4} takes a much higher value as compared to the peak on the negative $\Delta$-axis. This asymmetry arises due to the fact that as $J>g$, $\sin\left(\Phi_n/2\right)\approx 0$ while $\cos\left(\Phi_n/2\right)\approx 1$ which favors the $\ket{\lambda_-}\longrightarrow\ket{g,0}$ transition over the $\ket{\lambda_+}\longrightarrow\ket{g,0}$ transition (see Fig.~\ref{Fig2}). In the atomic excitation probability (Fig.~\ref{Fig4}(c)) we don't observe such an asymmetry as our setup is pumped with a light field directly shined onto the cavity mirrors that causes more impact on the cavity bare states.

Next, we turn our attention to the case when both cavity mirrors are shined with laser fields of equal phase and amplitude ($\hat{a}^{(l)}_{in}=\hat{a}^{(r)}_{in}$). As we have already discussed in Sec.~IV A if the CPA conditions given in Eq.~\eqref{CPAcond} are met, perfect photon trapping would occur. In Fig.~\ref{Fig5}(a), we numerically solve and plot the normalized output field intensities $\left|\left<\hat{a}_{out}\right>\right|^2/\left|\left<\hat{a}_{in}\right>\right|^2$ as a function of detuning $\Delta$. We confirmed the existence of CPA for both $N=1$ and $N=2$ with $J=0$ cases (gray dotted and thick solid red curves) where Eq.~\eqref{CPAcond} conditions are satisfied. In other words at the two frequencies where the photons are perfectly absorbed are the frequencies where the incoming laser is tuned with the polariton resonance $\pm \sqrt{2g^2n}$. Furthermore, the intracavity field intensity $\left|\left<\hat{a}\right>\right|^2/\left|\left<\hat{a}_{in}\right>\right|^2$ and collective atomic excitation probabilities $\left|\left<\hat{\sigma}_{1}+\hat{\sigma}_2\right>\right|^2/\left|\left<\hat{a}_{in}\right>\right|^2$ plotted in Fig.~\ref{Fig5}(b) and Fig.~\ref{Fig5}(c) shows that at the same frequencies the photon energy is completely absorbed within the CQED system. As a key finding, we notice that the enhancement in the DDI destroys the CPA (see blue dotted and brown thin solid curves in Fig.~\ref{Fig5}(a)) by creating a mismatch between the left and right-hand sides of conditions described in Eq.~\eqref{CPAcond}. Particularly, when the value of DDI exceeds the atom-cavity coupling g i.e. when $J>g$ this behavior, as evident from the brown thin solid curve, becomes most pronounced such that the CPA diminishes to almost $35\%$ with $J/g=3/2$. 

\begin{figure*}
\begin{tabular}{c c c}
\centering 
\includegraphics[width=2.2in, height=1.42in]{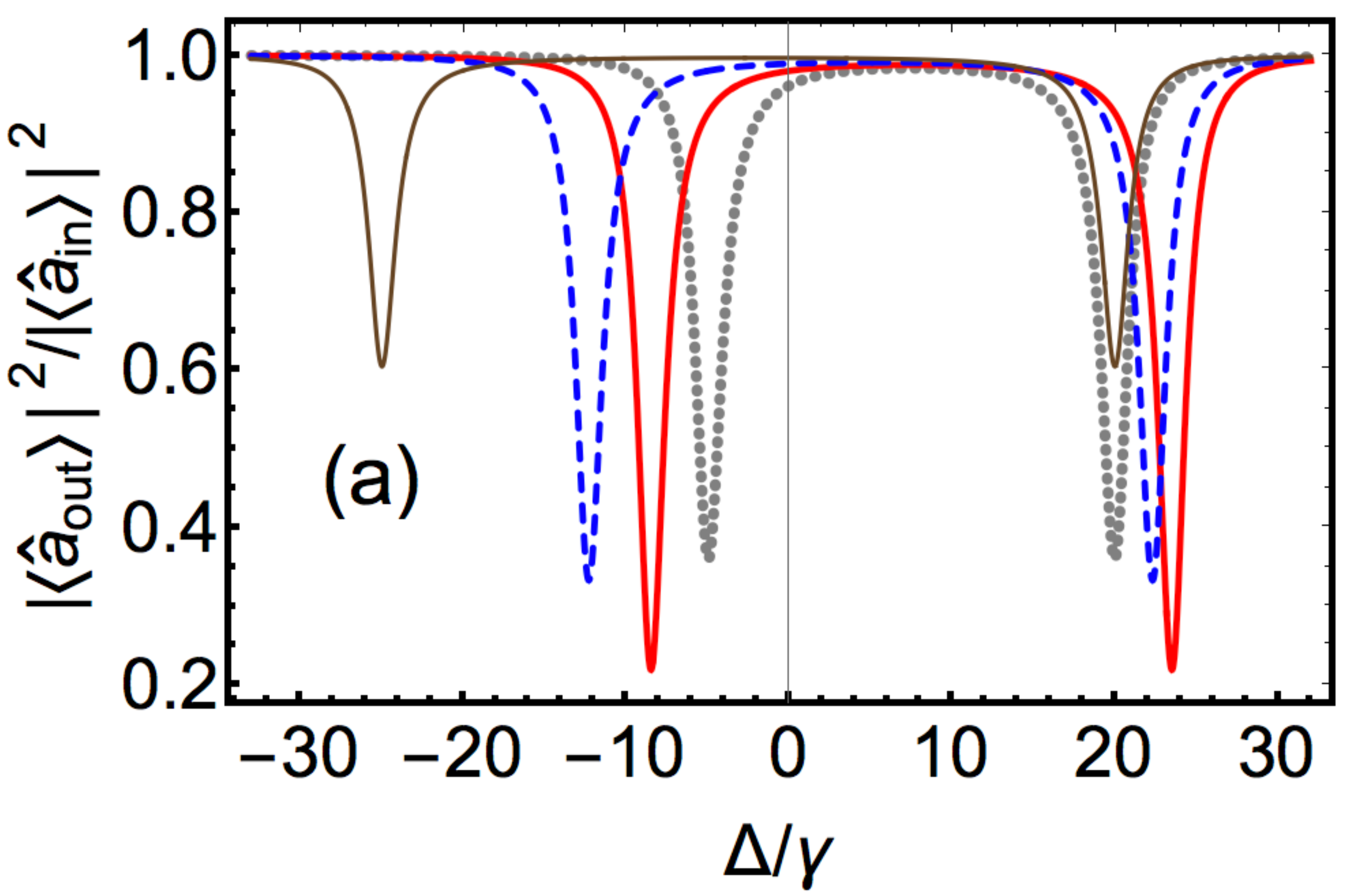} &
\includegraphics[width=2.2in, height=1.42in]{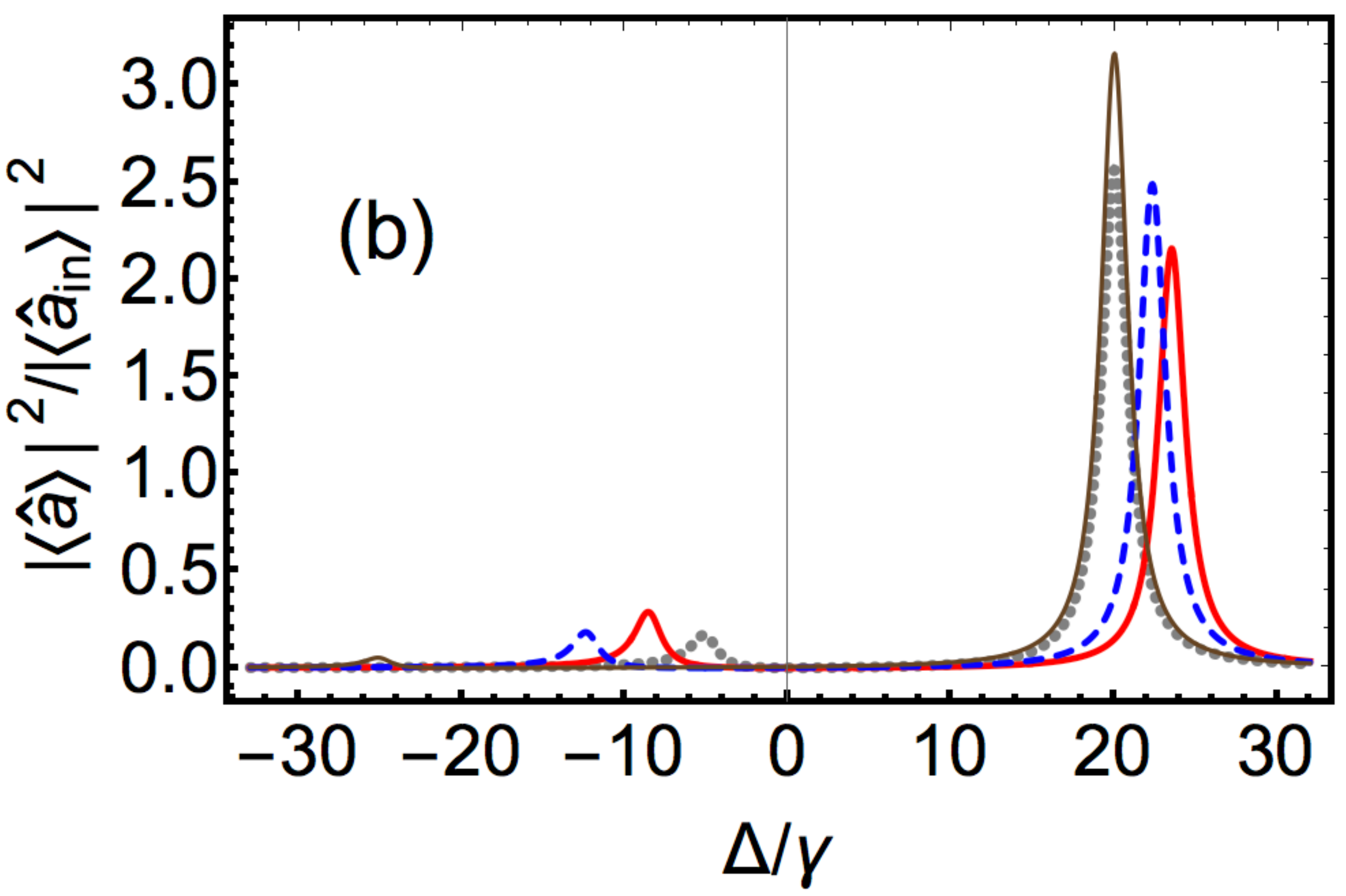} &
\includegraphics[width=2.2in, height=1.42in]{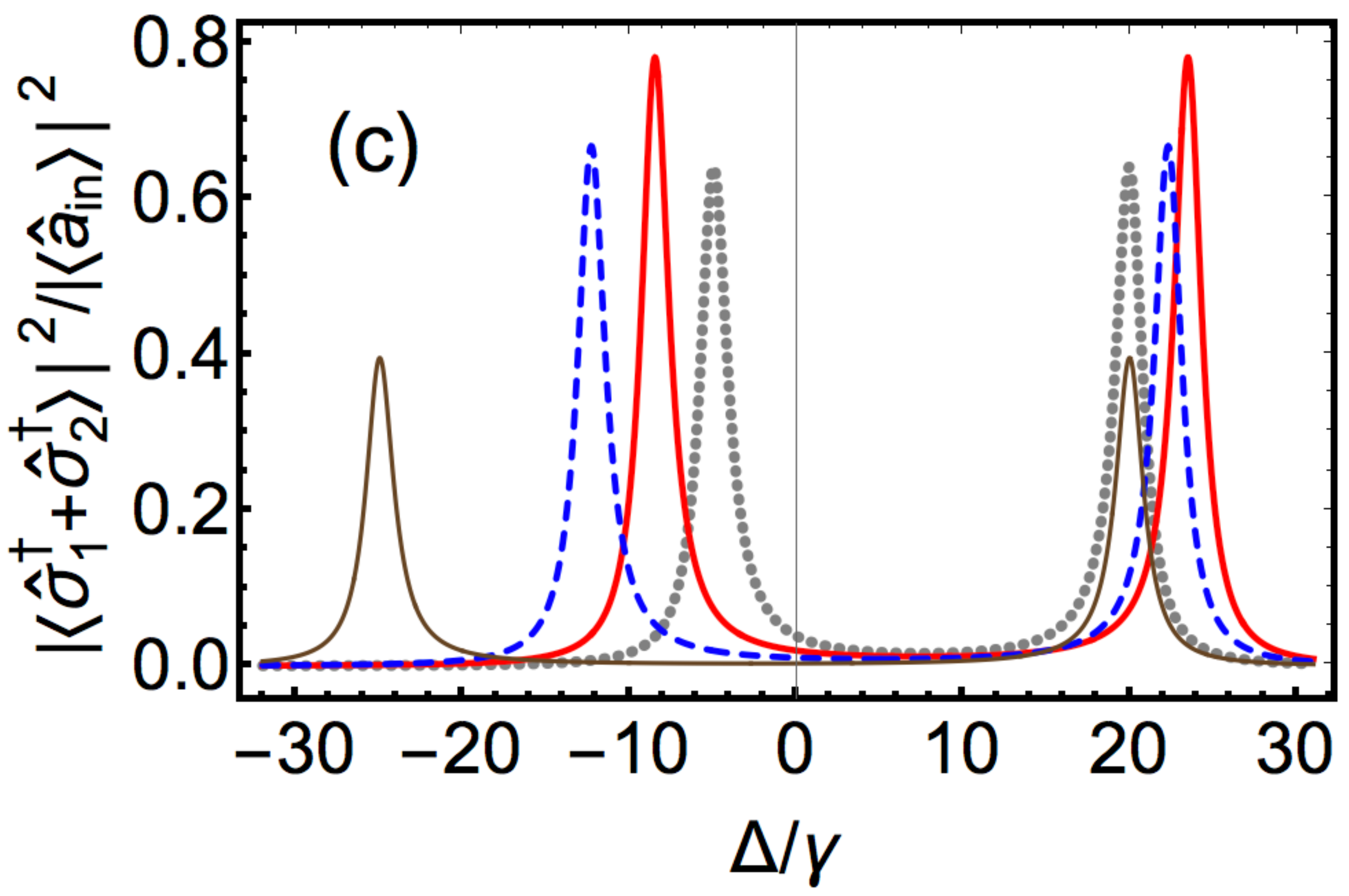} \\
\includegraphics[width=2.2in, height=1.42in]{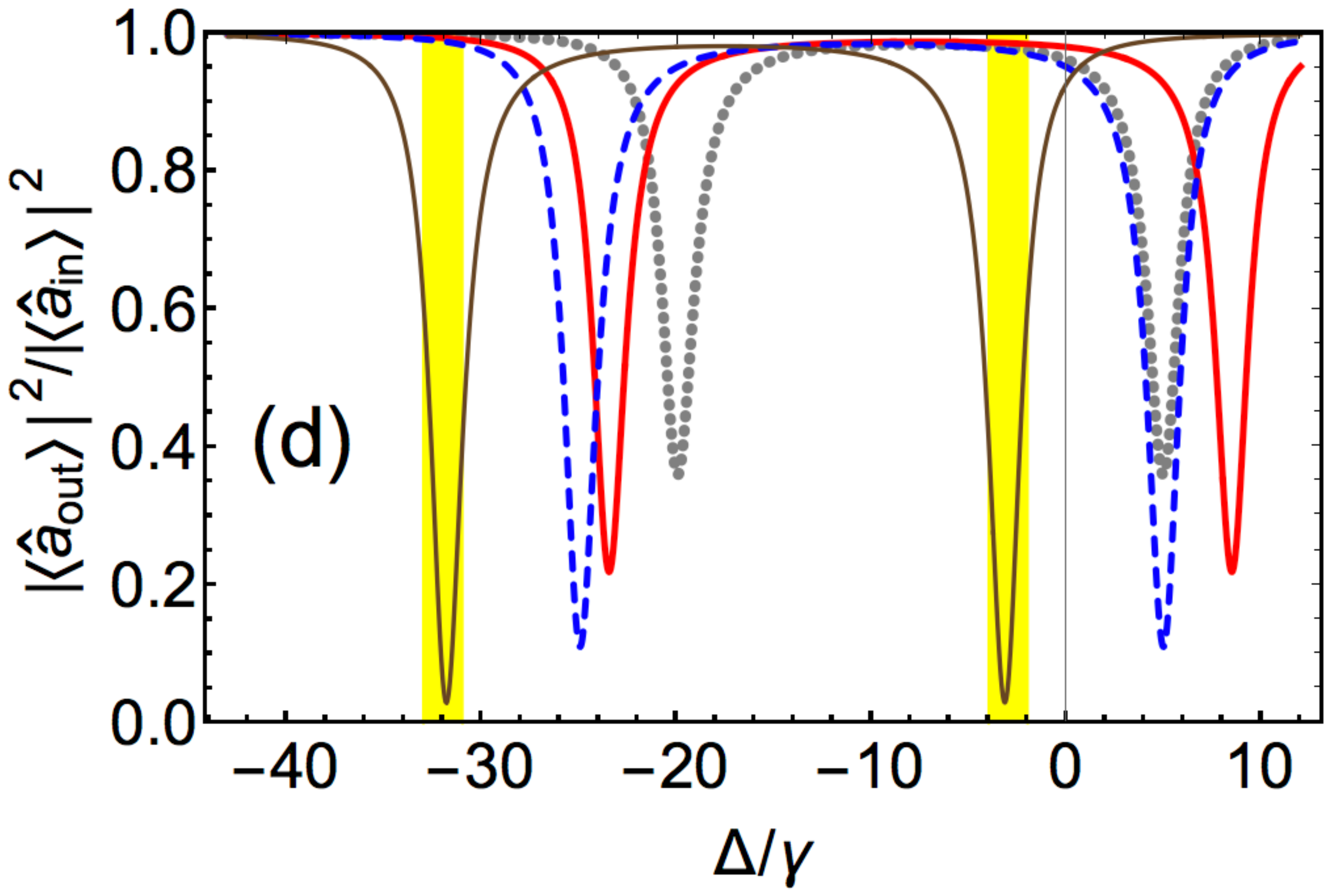} &
\includegraphics[width=2.2in, height=1.42in]{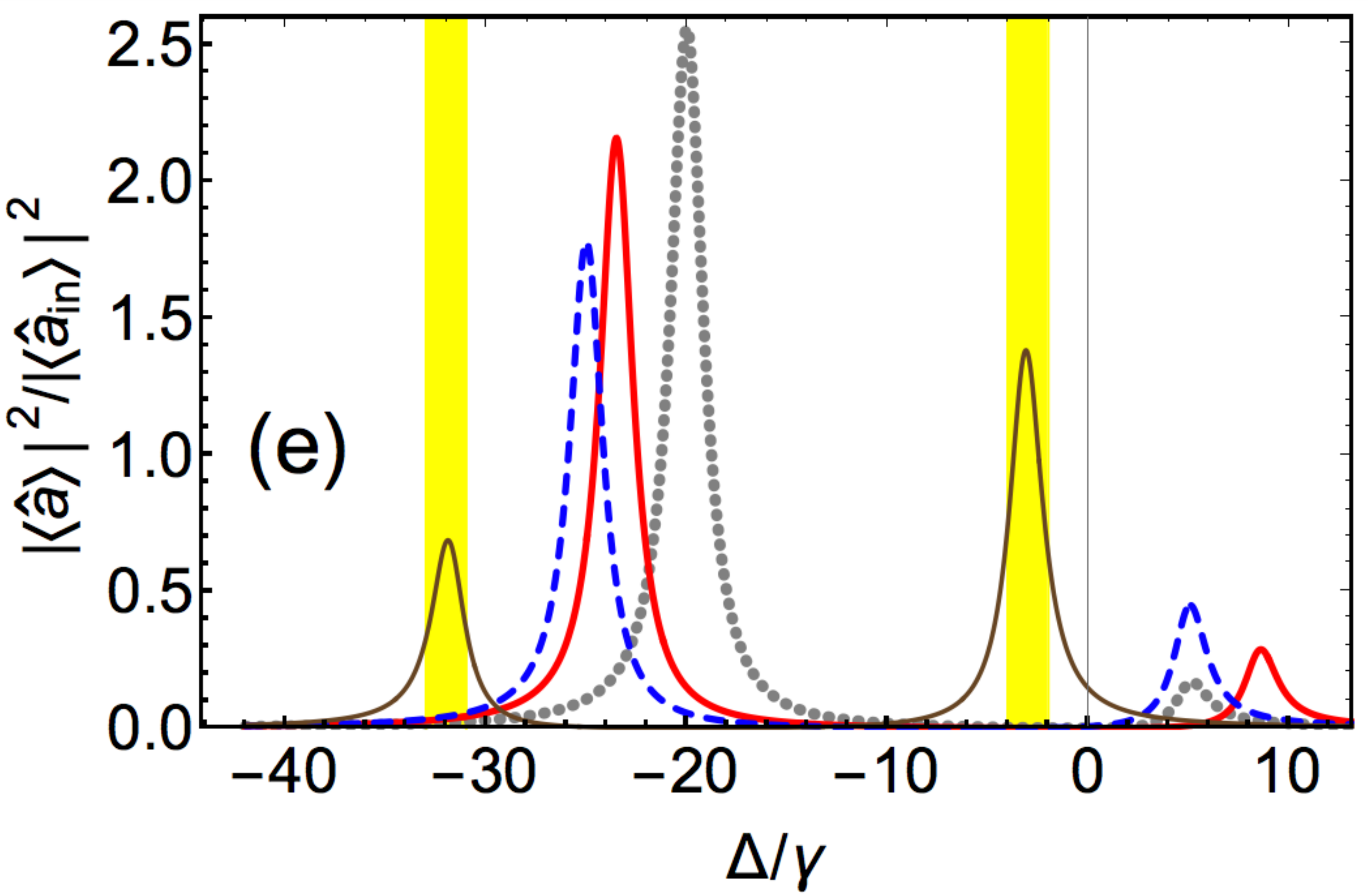} &
\includegraphics[width=2.2in, height=1.42in]{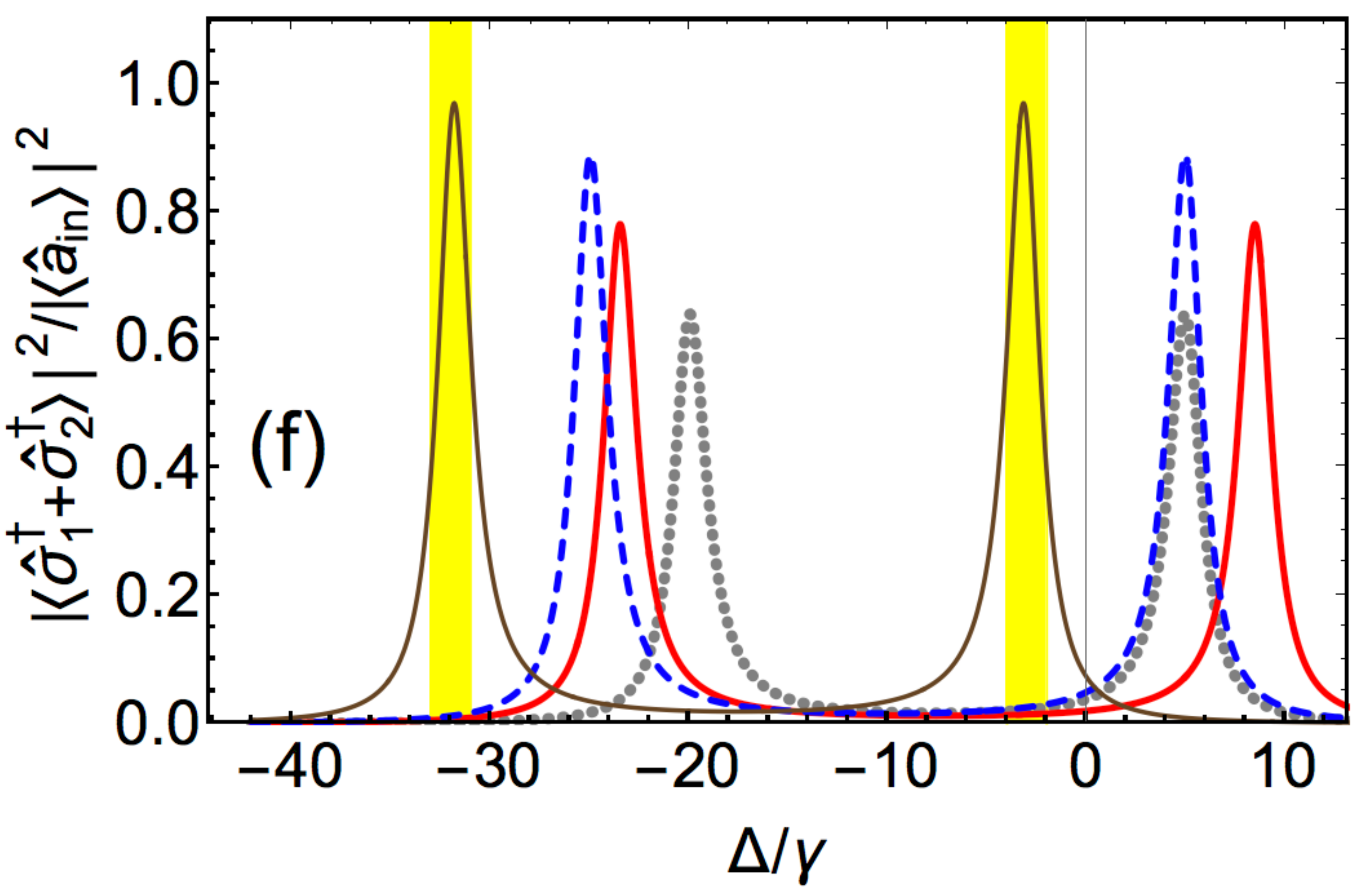} \\
\end{tabular}
\captionsetup{
format=plain,
margin=0.1em,
justification=raggedright,
singlelinecheck=false
}
\caption{Outfield intensities and collective atomic excitation probabilities for an off-resonance case. In the upper and lower rows of plots, we have considered $\Delta_{ac}=15\gamma$ and $\Delta_{ac}=-15\gamma$, respectively. Brown solid curves represent cases with $J=20\gamma$. The rest of the parameters are the same as used in Fig.~\ref{Fig4}. Yellow highlighted regions in the bottom row of plots show the frequency points of perfect photon absorption. Note that, due to the selection of a far-off resonance scenario, in some of the plots we have extended the range of $\Delta/\gamma$ values on the horizontal axis to capture all spectral features.}\label{Fig6}
\end{figure*}

In the results, up till now, we have assumed on-resonance conditions i.e. we have set $\omega_c=\omega_{eg}$. An important question in this context is in what ways an off-resonant situation can impact the CPA in the presence of DDI? We now address this question. From the CPA conditions mentioned in Eq.~\eqref{CPAcond}, we find four sets of possible values for detunings $\Delta_{eg}$ and $\Delta_c$ out of which two sets can take real values and are given by
\begin{align}\label{CPAonDel}
&\Delta_{eg}=-J\pm\sqrt{\frac{2g^2\gamma}{\kappa}-\gamma^2}, ~~~\Delta_c=\pm\sqrt{\frac{2g^2\kappa}{\gamma}-\kappa^2},\nonumber\\
&\implies \Delta_{ac} = -J\pm\sqrt{\frac{2g^2\gamma}{\kappa}-\gamma^2} \mp \sqrt{\frac{2g^2\kappa}{\gamma}-\kappa^2},
\end{align}
where the DDI introduces a shift in the atom-laser (and consequently atom-cavity) detuning frequency. Additionally, from the dressed state analysis as shown in Eq.~\eqref{Eigenpm}, we have found that the polariton state frequency is given by  $\lambda_{\pm}/\hbar=\frac{1}{2}\widetilde{\Delta}_{ac}\pm \frac{1}{2}\widetilde{\Omega}_n\left(\widetilde{\Delta}_{ac}\right)$. We observe that for the perfect photon trapping to occur both conditions needed to be satisfied simultaneously. The second condition on $\lambda_{\pm}$ would ensure the excitation of polariton states while the first condition (as mentioned in Eq.~\eqref{CPAonDel}) would then fulfill the requirement for a perfect absorption of photons interacting with the atom-cavity setup. In Fig.~\ref{Fig6} we plot the output and inter-cavity field intensities along with the collective atomic excitation probabilities subjected to an off-resonant situation (i.e when $\omega_c\neq\omega_{eg}$). In the upper (lower) row of plots, we take $\Delta=15\gamma$ ($\Delta=-15\gamma$) while in all plots a strong coupling regime of CQED has been assumed. In the $\Delta=15\gamma$ case we notice in Fig.~\ref{Fig6}(a) as DDI is elevated, the DDI and atom-cavity detuning act together to suppress the CPA. On the contrary, in Fig.~\ref{Fig6}(d) the negative detuning combined with the strong DDI tends to promote the CPA. As a result, for the specially chosen numerical parameters of Fig.~\ref{Fig6}, we observe the perfect photon trapping at $\Delta=-3\gamma$ and $\Delta=-32\gamma$. From Fig.~\ref{Fig6}(a), (d) and as well from reference \cite{agarwal2015photon}, we emphasize that the presence of CPA provides a better control (through changing the inter-atomic separation) and allowed us to achieve CPA at two tunable frequencies which are not possible to attain with a single QE case in the presence of detuning (see grey-dotted curves in Fig.~\ref{Fig6}(a) and (d)). As a last noticeable point in this figure, we note from Fig.~\ref{Fig6}(e) and Fig.~\ref{Fig6}(f) that at the CPA frequencies the contribution in the polariton state is dominated by the atomic excited state as compared to the cavity field contribution which is consistent with the arguments presented in Ref.~\cite{zanotto2014perfect, agarwal2014nanomechanical}.

\begin{figure*}
\begin{tabular}{c c c}
\centering 
\includegraphics[width=2.275in, height=1.65in]{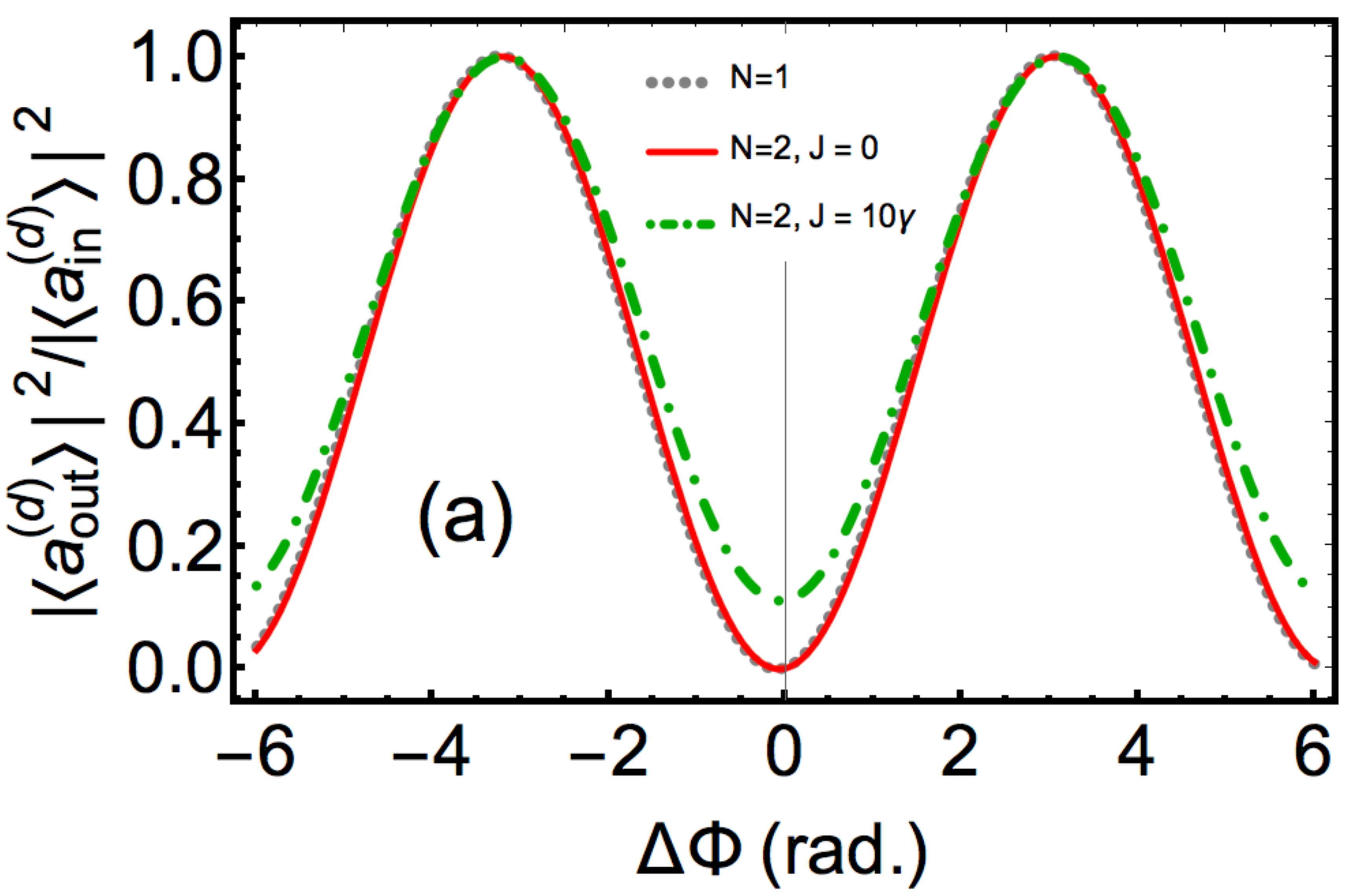} 
\includegraphics[width=2.275in, height=1.65in]{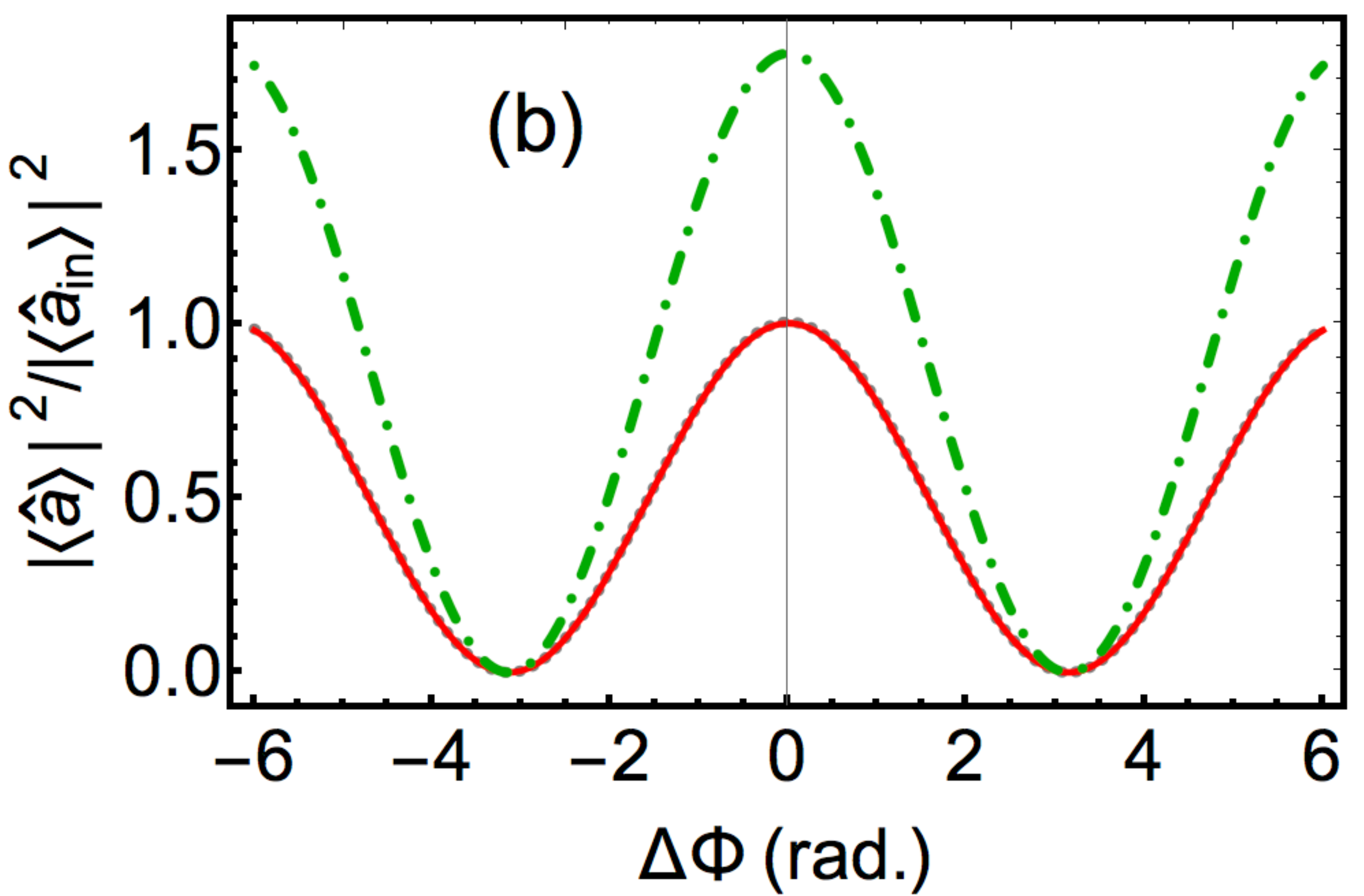} 
\includegraphics[width=2.275in, height=1.65in]{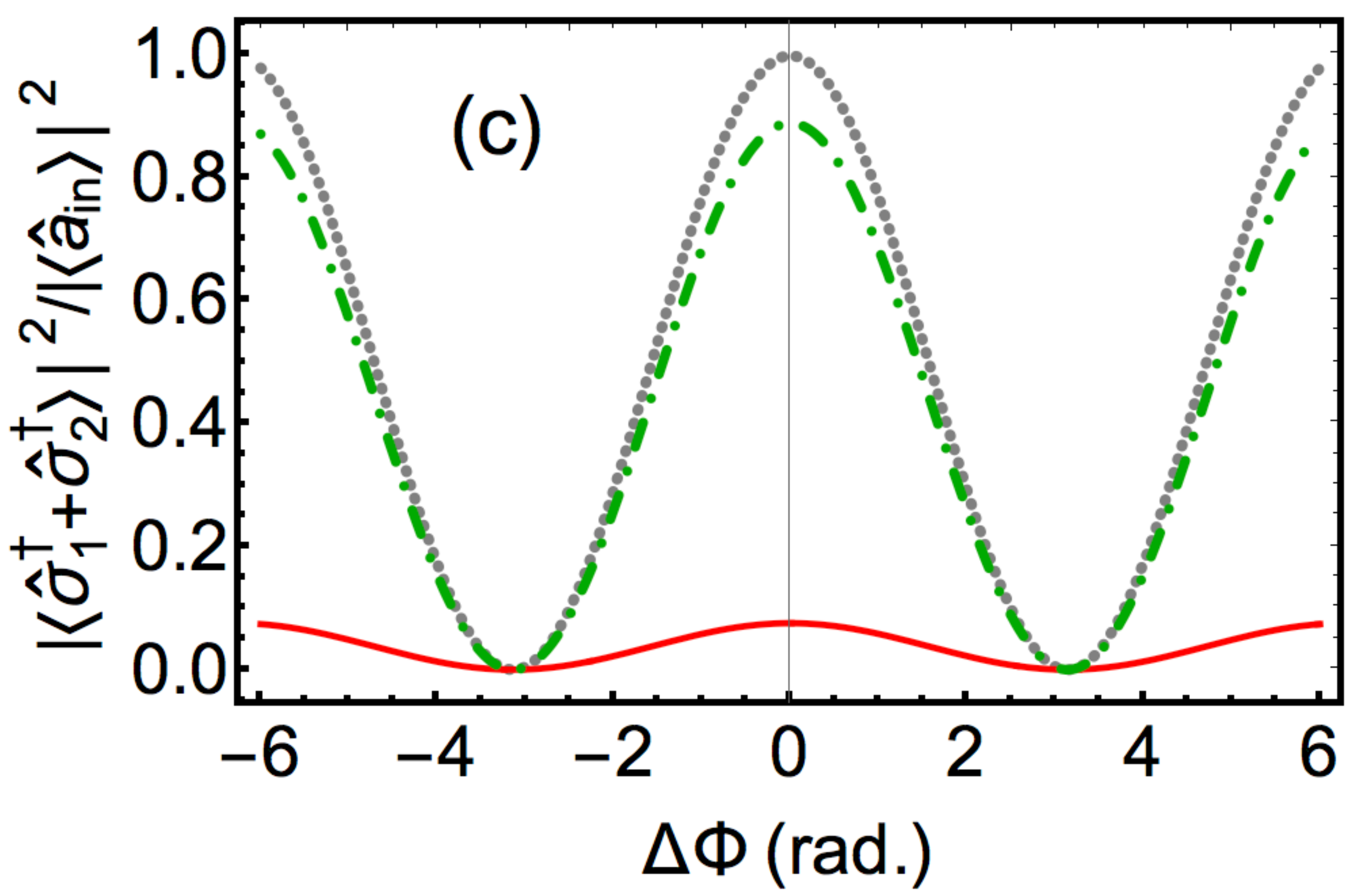}
\end{tabular}
\captionsetup{
format=plain,
margin=0.1em,
justification=raggedright,
singlelinecheck=false
}
\caption{Plots showing the dependence of normalized optical properties on the relative phase $\Delta\Phi=\varphi_{l}-\varphi_{r}$ between the two input fields for the cases of a single QE, two QEs without any DDI, and two QEs with strong DDI of $10\gamma$. {\bf(a)} output field intensity with $d = l,r$, {\bf(b)} intra-cavity field intensity, and {\bf(c)} atomic excitation probability. In plot {\bf(a)}, due to same choice of parameters for both atoms, the output intensities in left and right channel follow an identical pattern. In plot {\bf(b)} and {\bf(c)} we have used i.e. $\hat{a}^{(l)}_{in}=\hat{a}^{(r)}_{in}\equiv \hat{a}_{in}$. As before, the same parameters of strong cavity QED have been considered except that now we have set $\Delta_{ac}=10\gamma$.}\label{Fig7}
\end{figure*}

The output spectral properties of our Tavis-Cummings model also depend on the relative phase between the two input fields, which up to this point has been set equal to zero in all results for simplicity's sake. We now introduced this relative phase $\Delta\Phi$ through the relations
\begin{align}
\left<\hat{a}^{(l)}_{in}\right>= \left|\left<\hat{a}^{(l)}_{in}\right>\right|e^{i\varphi_l},~~~\text{and}~~~~\left|\left<\hat{a}^{(r)}_{in}\right>\right|e^{i\varphi_r}.
\end{align}
In Fig.~\ref{Fig7}, we report the spectral properties of our setup as we vary the relative phase $\Delta\Phi$ between $-2\pi$ to $2\pi$. For comparisons, we present the cases of a single QE (grey dotted curve), two QEs with no DDI (solid red curve), and two QEs with a strong DDI of $10\gamma$ (green dotted dashed curve). First of all, in all plots, we notice a symmetric profile around $\Delta\Phi=0$, while the output field intensities increases from $0$ to their maximum value of $1$ as $\Delta\Phi$ is varied from $0$ to $\pi$. Whereas, the intra-cavity field and atomic excitation probability change from their respective maximum values to the minimum value of $0$. This pattern extends down to all three cases of $N=1$; $N=2, J=0$; and $N=2, J=10\gamma$. We further point out that we observe a similar trend in our spectra for the on-resonance i.e. $\Delta_{ac}$ case as well (results not shown here). However, in the presence of a positive detuning of $10\gamma$, the differences between the $N=2, J\neq 0$ case (green dotted dashed curve), and other cases are most pronounced in the plot (b) and (c). In both plots (b) and (c), we observe that the positive detuning in some sense works in conjunction with the strong DDI to produce considerably elevated maximum values of atomic excitation probability and thereby intra-cavity field intensity as compared to the corresponding $N=2, J=0$ case, and thus can be utilized to distribute the incoming photon energy between the optical mode and the atomic degrees of freedom.


\section{Summary and Conclusions}
In this paper, we have studied the presence or absence of CPA of coherent light fields in the Tavis-Cummings models of CQED. We derived the quantum Langevin equations and input-output relations and studied the optical properties of the system under the low-excitation limit, mean-field approximation, and symmetric cavity assumption. We mathematically predicted the conditions required to achieve CPA in the presence of DDI and developed further physical understanding of the process by performing the dressed-state analysis involving the polariton states manifold. 

In our main findings, we concluded that a single input is not enough for the perfect destructive interference between the incoming laser field and the field emitted by our CQED system. And even in the presence of the two input fields, strong DDI tends to destroy the CPA when the QEs and cavity mode are in resonance. This behavior extended down to the scenario when atoms and fields were positively detuned. However, as the main result of this work, we found that some values of negative detuning (for instance $-15\gamma$) can cancel the impact of strong DDI on the CPA allowing the perfect absorption of the photons at two controllable laser frequencies under the strong coupling regime of CQED. We compared this behavior to a single emitter CQED case in the presence of detuning and found that this behavior is special to the Tavis-Cummings models (without the possibility to be achieved with the single QE CQED studies conducted earlier). 

Finally, we discussed the dependence of the CPA on the relative phase between the incoming light fields. We observed a periodic pattern of outfield intensities as a function of relative phase and concluded that a strong DDI and a positive detuning can generate considerable enhancement in the intra-cavity and atomic excitation peak heights even in the presence of incoherent dissipation from the polariton states. With the current progress in the CQED and circuit QED platforms, our results may find applications in the development of quantum photonic storage devices such as quantum memories with usage in building quantum communication networks \cite{lvovsky2009optical, kimble2008quantum}.


\acknowledgments
IMM would like to thank Michael J. Crescimanno for bringing the subject of CPA to our attention. This work is supported by the Miami University College of Arts and Science and Physics Department start-up funding.


\appendix

\section{Quantum Langevin Equations}
In this Appendix we outline the derivation of the quantum Lengevin equations \cite{steck2007quantum} which lay the foundations of our open quantum system treatment in the Heisenberg picture. To this end, we first notice that the relevant system operators in the present problem are: $\hat{a}(t)$, $\hat{\sigma}_{j}(t)$, $\hat{\sigma}_{z,j}(t)$; while baths are represented in terms of the operators $\hat{b}_{l}(\omega;t)$, $\hat{b}_{r}(\nu;t)$, and $\hat{S}_{j}(\chi_j;t)$ with $j=1,2$, $k=1,2$, $J_{jk}=J_{kj}$, and $\chi_1=\mu$, $\chi_2=\xi$. Note that in this notation $\omega, \nu,$ and $\chi_j$ are not frequencies conjugate to the time variable $t$ in the Fourier space sense, rather these are the frequencies labeling the bath modes for the cavity field and QEs. Using the Heisenberg equation of motion we find that these operators obey the following set of coupled differential equations
\begin{widetext}
\begin{subequations}\label{HEOM}
\begin{align}
&\partial_t\hat{a}(t) = -i\Delta_c\hat{a}(t)-i\sum\limits^{2}_{j=1}g^\ast_j\hat{\sigma}_j(t)-i\sqrt{\frac{\kappa_l}{2\pi}}\int^\infty_{-\infty}d\omega\hat{b}_l(\omega;t)-i\sqrt{\frac{\kappa_r}{2\pi}}\int^\infty_{-\infty}d\nu\hat{b}_r(\nu;t), \label{HEOM1}\\    
&\partial_t\hat{\sigma}_j(t) = -i\Delta_{eg_j}\hat{\sigma}_j(t)+ig_j\hat{\sigma}_{z,j}(t)\hat{a}(t)+iJ_{jk}\hat{\sigma}_{z,j}(t)\hat{\sigma}_k(t)+i\sqrt{\frac{\gamma_j}{\pi}}\hat{\sigma}_{z,j}(t)\int^\infty_{-\infty}d\mu\hat{S}_j(\mu;t), \label{HEOM2}\\    
&\partial_t\hat{\sigma}_{z,j}(t) = -2i\left(g_j\hat{a}(t)\hat{\sigma}^\dagger_j(t)-g^\ast_j\hat{\sigma}_j(t)\hat{a}^\dagger(t)\right)-2iJ_{jk}\left(\hat{\sigma}^\dagger_j(t)\hat{\sigma}_k(t)-\hat{\sigma}^\dagger_k(t)\hat{\sigma}_j(t)\right)\nonumber\\
&\hspace{18.5mm}-2i\sqrt{\frac{\gamma_j}{\pi}}\left(\hat{\sigma}^\dagger_j(t)\int^\infty_{-\infty}d\mu\hat{S}(\mu;t)-\int^\infty_{-\infty}d\mu\hat{S}^\dagger(\mu;t)\hat{\sigma}_j(t)\right),\label{HEOM3}\\   
&\partial_t\hat{b}_l(\omega;t) = -i\omega\hat{b}_l(\omega;t)-i\sqrt{\frac{\kappa_l}{2\pi}}\hat{a}(t),\label{HEOM4}\\
&\partial_t\hat{b}_r(\nu;t) = -i\omega\hat{b}_r(\nu;t)-i\sqrt{\frac{\kappa_r}{2\pi}}\hat{a}(t),\label{HEOM5}\\
&\partial_t\hat{S}_j(\chi_j;t) = -i\chi_j\hat{S}_j(\chi_j;t)-i\sqrt{\frac{\gamma_j}{\pi}}\hat{\sigma}_j(t). \label{HEOM6}
\end{align}
\end{subequations}
\end{widetext}
We remark that the structure of the above equation set allows us to straightforwardly extend the problem to $N$ number of QEs, in which case one would need to solve a set of $3(N+1)$ coupled differential equations. For the present case, proceeding with the derivation as the next step, we eliminate the bath operators from Eq.~\eqref{HEOM1}, \eqref{HEOM2}, and \eqref{HEOM3} and solve for the system dynamics. For this purpose, we take Eq.~\eqref{HEOM4} and apply the transformation: $\hat{b}_l(\omega;t)\longrightarrow \hat{b}_l(\omega;t)e^{i\omega t}$ under which we find
\begin{equation}\label{inOutMain}
\partial_t\left[\hat{b}_l(\omega;t)e^{i\omega t}\right]=-i\sqrt{\frac{\kappa_l}{2\pi}}\hat{a}(t)e^{i\omega t}.
\end{equation}
Next, we integrate the above equation from some arbitrary past time $t_0$ to some present time $t$ and obtain the solution
\begin{equation}
\hat{b}_l(\omega;t)=\hat{b}_l(\omega; t_0)e^{-i\omega(t-t_0)}-i\sqrt{\frac{\kappa_l}{2\pi}}\int^t_{t_0}dt^{'}\hat{a}(t^{'})e^{-i\omega (t-t^{'})}.\label{HEOM4a}
\end{equation}
Following the same line of calculations, we find that the other bath operators obey
\begin{align}
\hat{b}_r(\nu;t)&=\hat{b}_r(\nu; t_0)e^{-i\nu(t-t_0)}-i\sqrt{\frac{\kappa_r}{2\pi}}\int^t_{t_0}dt^{'}\hat{a}(t^{'})e^{-i\nu (t-t^{'})},\label{HEOM4b}\\
\hat{S}_j(\chi_j;t)&=\hat{S}_j(\chi_j;t_0)e^{-i\chi_j(t-t_0)}-i\sqrt{\frac{\gamma_j}{\pi}}\int^t_{t_0}dt^{'}\hat{\sigma}_j(t^{'})\nonumber\\
&\times e^{-i\mu (t-t^{'})}.\label{HEOM4c}
\end{align}
Inserting Eq.~\eqref{HEOM4a}, \eqref{HEOM4b}, and \eqref{HEOM4c} into Eq.~\eqref{HEOM1}, \eqref{HEOM2}, and \eqref{HEOM3} (with $\hat{b}_l(\omega;t_0)$, $\hat{b}_r(\nu;t_0)$, and $\hat{S}_j(\chi_j;t_0)$ essentially entering as the initial conditions for the left cavity channel, right cavity channel and QEs) and defining the set of input operators \cite{gardiner1985input} as
\begin{subequations}\label{InOp}
\begin{align}
\int^\infty_{-\infty}d\omega\hat{b}_l(\omega;t_0)e^{-i\omega(t-t_0)}d\omega &:=-i\sqrt{2\pi}\hat{a}^{(l)}_{in}(t),\\
\int^\infty_{-\infty}d\nu\hat{b}_r(\nu;t_0)e^{-i\nu(t-t_0)} &:=-i\sqrt{2\pi}\hat{a}^{(r)}_{in}(t),\\
\int^\infty_{-\infty}d\chi_j\hat{S}_j(\chi_j;t_0)e^{-i\chi_j(t-t_0)} &:=-i\sqrt{2\pi}\hat{\sigma}_{in,j}(t),
\end{align}
\end{subequations}
we arrive at the following set of quantum Langevin equations describing the time evolution of system operators $\hat{a}$, $\hat{\sigma}_j$ and $\hat{\sigma}_{z,j}$ 
\begin{subequations}\label{HEOM}
\begin{align}
\partial_t\hat{a}(t) =& -\Big(i\Delta_c+\frac{\kappa_l+\kappa_r}{2}\Big)\hat{a}(t)-i\sum\limits^{2}_{j=1}g^\ast_j\hat{\sigma}_j(t)\nonumber\\
&-\sqrt{\kappa_l}\hat{a}^{(l)}_{in}(t)-\sqrt{\kappa_r}\hat{a}^{(r)}_{in}(t), \label{QLE1}\\    
\partial_t\hat{\sigma}_j(t) =& -\left(i\Delta_{eg_j}+\gamma_j\right)\hat{\sigma}_j(t)+ig_j\hat{\sigma}_{z,j}(t)\hat{a}(t)\nonumber\\
&+iJ_{jk}\hat{\sigma}_{z,j}(t)\hat{\sigma}_k(t)+\sqrt{2\gamma_j}\hat{\sigma}_{z,j}(t)\hat{\sigma}_{in,j}(t), \label{QLE2}\\    
\partial_t\hat{\sigma}_{z,j}(t) =& -4\gamma_j\Big(\frac{\hat{\sigma}_{z,j}(t)+\hat{1}}{2}\Big)
-2i\Big(g_j\hat{a}(t)\hat{\sigma}^\dagger_j(t)\nonumber\\
&-g^\ast_j\hat{\sigma}_j(t)\hat{a}^\dagger(t)\Big)
-2iJ_{jk}\Big(\hat{\sigma}^\dagger_j(t)\hat{\sigma}_k(t)\nonumber\\
&-\hat{\sigma}^\dagger_k(t)\hat{\sigma}_k(t)\Big)-2\sqrt{2\gamma_j}\Big(\hat{\sigma}^\dagger_j(t)\hat{\sigma}_{in,j}(t)\nonumber\\
&+\hat{\sigma}^{\dagger}_{in,j}(t)\hat{\sigma}_j(t)\big).\label{QLE3}
\end{align}
\end{subequations}
We notice that there are three types of terms that are arising in the above set of Langevin equations. The terms with detunings and atom-cavity coupling rates describe the internal dynamics of the system. The terms with prefactors $\kappa_l,\kappa_r$, and $\gamma_j$ are the decay terms representing the processes of energy loss from the system to the environment. Finally, the terms with prefactors $\sqrt{\kappa_l}$, $\sqrt{\kappa_r}$, and $\sqrt{\gamma_j}$ are the quantum noise/input terms describing the impact of baths on the system at some past time $t_0$.


\section{Input-Output relations}
We begin from Eq.~\eqref{inOutMain} and (as before) we integrate it from some past time $t_0$ to some present time $t$. We then integrate the resultant equation over all possible environment modes (using $\int^{+\infty}_{-\infty}d\omega e^{i\omega (t-t^{'})}=2\pi\delta(t-t^{'})$) and by recognizing the input operator we arrive at the following equation for the left channel
\begin{align}\label{pastbl}
\int^\infty_{-\infty}d\omega\hat{b}_l\left(\omega;t \right) = -i\sqrt{2\pi}\hat{a}^{(l)}_{in}(t)-i\sqrt{\frac{\pi\kappa_l}{2}}\hat{a}(t).
\end{align}
Next, we specify an arbitrary future time $t_1$ and now integrate the Eq.~\eqref{inOutMain} from present time $t$ to $t_1$. After integrating over all possible $\omega$ values we find
\begin{align}\label{futurebl}
\int^\infty_{-\infty}d\omega\hat{b}_l\left(\omega;t \right) = -i\sqrt{2\pi}\hat{a}^{(l)}_{out}(t)+i\sqrt{\frac{\pi\kappa_l}{2}}\hat{a}(t),
\end{align}
where we identify the output operator for the field propagating in the left channel $\hat{a}^{(l)}_{out}(t)$ as 
\begin{align}
-i\sqrt{2\pi}\hat{a}^{(l)}_{out}(t):=\int^{\infty}_{-\infty} d\omega \hat{b}_{l}\left(\omega;t_1\right)e^{-i\omega(t_1-t)}d\omega.
\end{align}
Comparing Eq.~\eqref{pastbl} and Eq.~\eqref{futurebl} we arrive at the following expression that relates the output field and input fields (commonly known as the Collett and Gardiner input-output relation \cite{gardiner1985input}) for the left channel
\begin{align}\label{inoutL}
\hat{a}^{(l)}_{out}(t)=\hat{a}^{(l)}_{in}(t)+\sqrt{\kappa_l}~\hat{a}(t).
\end{align}
Following similar lines of calculations, we find the following set of input-output relations for the right cavity channel and QEs
\begin{subequations}\label{InOutAll}
\begin{align}
\hat{a}^{(r)}_{out}(t)&=\hat{a}^{(r)}_{in}(t)+\sqrt{\kappa_r}~\hat{a}(t),\\
\hat{\sigma}_{out,j}(t)&=\hat{\sigma}_{in,j}(t)+\sqrt{\gamma_j}~\hat{\sigma}_j(t).
\end{align}
\end{subequations}
Here $\hat{a}^{(r)}_{out}(t)$ and $\hat{a}^{(r)}_{in}(t)$ are the output and input field operators for the right channel, respectively while $\hat{\sigma}_{out,j}(t)$ and $\hat{\sigma}_{in,j}(t)$ are the respective output and input operators for the $j$th QE. These output operators are defined as
\begin{subequations}\label{OutRS}
\begin{align}
-i\sqrt{2\pi}\hat{a}^{(r)}_{out}(t):&=\int^{\infty}_{-\infty} d\nu \hat{b}_{r}\left(\nu;t_1\right)e^{-i\nu(t_1-t)}d\nu,\\
-i\sqrt{2\pi}\hat{\sigma}_{out,j}(t):&=\int^{\infty}_{-\infty} d\chi_j \hat{S}_{j}\left(\chi_j;t_1\right)e^{-i\chi_j(t_1-t)}d\chi_j.
\end{align}
\end{subequations}
We make this remark that the generic structure of the aforementioned input-output relations essentially works as boundary conditions relating the output field with the input fields while incorporating the system response. We also note that the construction of the input, output, and system operators in the input-output formalism is such that the condition of causality is always satisfied \cite{gardiner2004quantum}.

\bibliographystyle{ieeetr}
\bibliography{paper.bib}
\end{document}